\documentclass{ieeeaccess}
\usepackage{cite}
\usepackage{amsmath,amssymb,amsfonts}
\usepackage{algorithmic}
\usepackage{array,booktabs}
\usepackage{xurl}
\usepackage{graphicx}
\usepackage{textcomp}

\usepackage{bm}
\makeatletter
\AtBeginDocument{\DeclareMathVersion{bold}
\SetSymbolFont{operators}{bold}{T1}{times}{b}{n}
\SetSymbolFont{NewLetters}{bold}{T1}{times}{b}{it}
\SetMathAlphabet{\mathrm}{bold}{T1}{times}{b}{n}
\SetMathAlphabet{\mathit}{bold}{T1}{times}{b}{it}
\SetMathAlphabet{\mathbf}{bold}{T1}{times}{b}{n}
\SetMathAlphabet{\mathtt}{bold}{OT1}{pcr}{b}{n}
\SetSymbolFont{symbols}{bold}{OMS}{cmsy}{b}{n}
\renewcommand\boldmath{\@nomath\boldmath\mathversion{bold}}}
\makeatother

\def\BibTeX{{\rm B\kern-.05em{\sc i\kern-.025em b}\kern-.08em
    T\kern-.1667em\lower.7ex\hbox{E}\kern-.125emX}}

\begin{document}
\bstctlcite{IEEEtranBSTcontrol}
\history{Date of publication xxxx 00, 0000, date of current version xxxx 00, 0000.}
\doi{10.1109/ACCESS.2024.0429000}

\title{Paired-Mixture Learning and Embedding Correction for Metadata-Free Anomalous Sound Detection}
\author{\uppercase{Seunghyeon Shin}\authorrefmark{1} and
\uppercase{Seokjin Lee}\authorrefmark{1,2}, \IEEEmembership{Member, IEEE}}
\address[1]{School of Electronic and Electrical Engineering, Kyungpook National University, Daegu 41566, Republic of Korea (e-mail: sh.shin@knu.ac.kr)}
\address[2]{School of Electronics Engineering, Kyungpook National University, Daegu 41566, Republic of Korea}
\tfootnote{This work was supported by the National Research Foundation of Korea (NRF) grant funded by the Korea government (MSIT) under Grant RS-2025-23523113. This work forms part of Seunghyeon Shin's doctoral dissertation at Kyungpook National University.}
\markboth{S. Shin and S. Lee: Paired-Mixture Learning and Embedding Correction for Metadata-Free Anomalous Sound Detection}
{S. Shin and S. Lee: Paired-Mixture Learning and Embedding Correction for Metadata-Free Anomalous Sound Detection}
\corresp{Corresponding author: Seokjin Lee (e-mail: sjlee6@knu.ac.kr).}

\begin{abstract}
Acoustic monitoring often relies on normal machine recordings with limited information about operating conditions. We first train a separator to estimate a normal target recording from a mixture of machine sounds. With the separator frozen, we train regional attention on pairs of mixtures that share the target recording and contain different interfering sounds. Differences between the paired embeddings provide the covariance estimate for correction before cosine nearest-neighbor scoring. We evaluate the method on the development-test recordings of seven machine types in the 2022 Detection and Classification of Acoustic Scenes and Events challenge. We summarize detection performance by the harmonic mean of source-domain, target-domain, and partial areas under the receiver operating characteristic curve. On the original recordings, a separator trained without external pre-training achieved a three-run mean of 64.56\%, close to the 64.87\% obtained by distillation from a pre-trained audio transformer. Using the same transformer as a frozen encoder gave a mean of 64.88\%. With competing machine sounds added to the test recordings, both implementations outperformed distillation at all seven tested interference levels.
\end{abstract}


\titlepgskip=-21pt

\maketitle

\section{Introduction}
\label{sec:introduction}

\PARstart{A}{coustic} sensing enables non-contact machine condition monitoring through recorded operating sounds. A microphone monitoring one machine may also capture sounds from nearby equipment~\cite{lee2024activityguided}. Anomalous sound detection (ASD) assesses whether a recording is consistent with normal operation. Failures occur infrequently, and their acoustic characteristics vary with the fault and operating conditions, making representative anomalous recordings difficult to collect~\cite{koizumi2019unsupervised,koizumi2019toyadmos}. These constraints motivate learning from normal recordings. Changes in operating speed, load, microphone arrangement, and ambient noise also alter normal sounds, so a detector must accommodate normal acoustic variation while preserving information about machine condition~\cite{harada2021toyadmos2,dohi2022mimiidg}.

Detailed operating metadata offer useful supervision. Machine identities and attributes such as speed or load can define auxiliary classes, allowing a model to learn discriminative features from normal recordings. Obtaining these annotations can require additional monitoring~\cite{fujimura2025improvements}, and their availability varies across machine types~\cite{nishida2024task2overview}. A recording collection may thus contain normal sounds from a known machine type without consistent descriptions of each recording's acquisition conditions. Label granularity also determines which distinctions training emphasizes and affects the resulting representation~\cite{wilkinghoff2024angular}.

The 2023 first-shot setting of the Detection and Classification of Acoustic Scenes and Events (DCASE) challenge evaluated systems on new machine types using their normal recordings and settings developed on other types~\cite{harada2023firstshot}. DCASE 2024 provided operating-condition attributes for only a subset of machine types~\cite{nishida2024task2overview}. Audio-derived pseudo-labels offer one source of supervision when operating annotations are unavailable~\cite{fujimura2025improvements}.

In this study, \emph{metadata-free} describes detection from normal recordings of a known machine type, without detailed operating annotations. Machine type identifies the target recording pool. The remaining normal recordings form a common pool of interference sounds. Once the target set is selected, the proposed networks take acoustic inputs and learn from the shared target waveform and the relationships between mixtures. Instance identifiers, operating attributes, and section or domain labels are excluded from training and scoring. We evaluate this setting across seven machine types and both domains of DCASE 2022 Task~2~\cite{dohi2022task2overview}.

We form two mixtures by combining the same normal target recording with different sounds from other machines. The target waveform supervises a separator trained on the normal recordings. Once the separator is frozen, the shared target provides supervision for regional aggregation, and the differences between paired embeddings provide the covariance estimate for correction.

Prior work averaged intermediate separation features and concatenated them across stages for ASD~\cite{shin2024representational}. Here, attention pools features within time--frequency regions before fusion across stages. We also apply the same paired-mixture training and correction to other frozen encoders. At inference, cosine distance to the nearest normal reference gives the score for each encoded unit. To assess robustness, we vary the level of sound added to the same test recording, which retains its original environmental sounds.

On the original recordings, the separator trained without external pre-training achieved a three-run mean close to that of Dis-multi-PL~\cite{fujimura2026pseudolabel}. Using Bidirectional Encoder representation from Audio Transformers (BEATs) iter3~\cite{chen2023beats}, the encoder that supplies the features for the Dis-multi-PL pseudo-labels, regional aggregation with embedding correction scored higher in the representative run and achieved a similar three-run mean. With other-machine sounds added, both proposed implementations outperformed Dis-multi-PL at all seven tested signal-to-noise ratios (SNRs).

The study makes three contributions.
\begin{enumerate}
\item Pairs of mixtures sharing a normal target recording provide supervision for regional aggregation. Contrastive learning aligns the two mixture views, and a consistency loss aligns each view with the target reference. We examine this readout on intermediate separation features and on representations learned through reconstruction, classification, self-supervision, and external pre-training.
\item Embedding correction combines normal-reference centering with covariance estimated from paired-mixture differences. We compare this estimate with covariance from original normal recordings while holding the separator, learned aggregation, and scorer fixed.
\item We evaluate how competing machine sounds and Gaussian white noise affect detection across seven SNRs with the trained systems and normal references held fixed.
\end{enumerate}

\section{Related work}
\label{sec:related_work}

\subsection{Learning from normal data}

ASD systems trained on normal recordings learn normal acoustic characteristics or relationships among normal samples. Anomaly scores are derived from reconstruction errors, estimated densities, or comparisons between learned representations and normal references~\cite{dohi2022task2overview,fujimura2025asdkit}. Such training may also incorporate auxiliary labels for machine identity or operating conditions.

The DCASE autoencoder baseline learns to reconstruct log-mel features from normal recordings. Its anomaly score is the mean squared error (MSE) of the reconstruction~\cite{dcase2022task2}. Normalizing flows estimate normal-data likelihoods through invertible transformations. Dohi et al. combined density estimation with a discriminative objective that assigns higher likelihoods to the target machine than to other instances of the same type~\cite{dohi2021flow}.

Auxiliary classification distinguishes machine types, machine IDs, sections, or operating conditions. Features before the classification layer can subsequently support normal-reference scoring. Angular-margin objectives such as ArcFace promote intra-class compactness and inter-class separation~\cite{deng2019arcface,wilkinghoff2024angular}. AdaProj accommodates within-class variability through class-specific subspaces~\cite{wilkinghoff2024adaproj}. For anomaly detection within a machine type, these features must also retain information about acoustic differences among recordings of that type.

\subsection{Self-supervision and pre-trained representations}

Self-supervised learning derives relationships directly from the recordings. Contrastive methods typically treat two augmented views of the same recording as a positive pair and other recordings as negatives. This approach has been applied to ASD using audio augmentation~\cite{shin2022contrastive}. SimSiam aligns paired views using projection and prediction networks with stop-gradient targets~\cite{chen2021simsiam}. Other contrastive approaches define training relationships using machine IDs~\cite{guan2023machineid}.

Audio encoders trained on larger datasets provide another source of features. BEATs  predicts acoustic-token targets~\cite{chen2023beats}. The Efficient Audio Transformer (EAT) uses teacher--student training with frame- and recording-level targets~\cite{chen2024eat}. Consistent Ensemble Distillation (CED) transfers audio-tagging knowledge from teacher ensembles~\cite{dinkel2024ced}. ASDKit evaluates frozen and fine-tuned BEATs and EAT features for ASD~\cite{fujimura2025asdkit}. Fujimura et al. formed positive pairs by adding sounds from other machines, then clustered the features learned by triplet training to generate pseudo-labels~\cite{fujimura2025improvements}. Fujimura and Toda trained a student detector using pseudo-labels obtained by clustering features from a pre-trained audio encoder~\cite{fujimura2026pseudolabel}. In our method, the paired mixtures train regional aggregation directly and provide the differences used to estimate the correction covariance. These approaches provide supervision when detailed operating labels are unavailable. Audio-derived clusters can also reflect background noise~\cite{fujimura2025improvements,fujimura2026pseudolabel}.

\subsection{Separation-based anomalous sound detection}

Separation can help distinguish a monitored machine from its acoustic surroundings. Shimonishi et al. extracted features from separated signals for an autoencoder. They also compared estimates from separation by machine ID and by machine type for anomaly scoring~\cite{shimonishi2023separation}. Activity-guided detection uses machine activity information to guide separation and mask the downstream autoencoder input and reconstruction error~\cite{lee2024activityguided}. Target signal enhancement has also been used to prepare training signals and pseudo-labels for discriminative ASD, using normal recordings together with auxiliary clean machine-sound or noise recordings~\cite{fujimura2025discriminative}.

The earlier intermediate-feature method in~\cite{shin2024representational} mixed normal target and non-target recordings and trained a separator to estimate the non-target component. That system pooled features from the encoder and Conformer blocks, concatenated the stage vectors, and scored them by Mahalanobis distance. Our implementation trains the separator to estimate the target component and learns regional aggregation from mixtures sharing the original target recording. Paired-mixture embedding differences provide the interference covariance for correction before cosine nearest-reference scoring, linking the comparison statistics to the acoustic relationship used for representation learning.

\subsection{Feature aggregation and representation correction}

Global average pooling (GAP) averages each feature channel across time and frequency, while learned pooling can vary the contributions of individual positions. Okabe et al. learned frame weights for attentive statistics pooling~\cite{okabe2018attentive}. Jiang et al. combined learned layer weights with frequency-wise multi-query attention pooling for ASD and fine-tuned the encoder using metadata-derived labels~\cite{jiang2025fmqap}. Our readout retains a separate summary for each predefined time--frequency region, preserving coarse location information before fusion. Paired mixtures supervise the aggregation of frozen acoustic features in both the separator and the encoder experiments.

Within-class covariance normalization scales feature directions using within-speaker covariance in speaker recognition~\cite{hatch2006wccn}. Local density-based score normalization adjusts input-to-reference distances according to the spacing around each normal reference~\cite{wilkinghoff2026local}. The proposed correction estimates covariance from embedding differences between mixtures sharing the same target segment. These statistics describe variation associated with changed non-target sounds and complement the normal references used for anomaly scoring.

\section{Proposed method}
\label{sec:method}

\subsection{Problem setting and framework}
\label{sec:framework}

\begin{figure*}[!t]
\centering
\includegraphics[width=.83\textwidth]{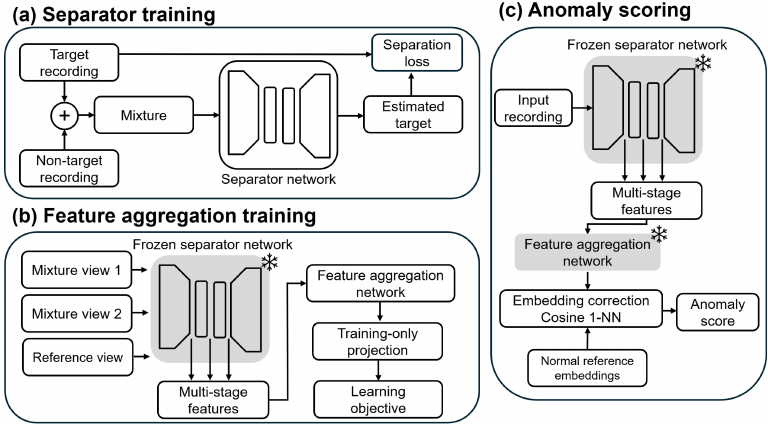}
\caption{Overview of the separation-based detector. A separator learns to estimate a normal target recording from a mixture. Paired mixtures of the same target then supervise feature aggregation on the frozen separator. At inference, the detector compares corrected embeddings with normal references for the monitored machine type.}
\label{fig:framework}
\end{figure*}

Fig.~\ref{fig:framework} shows the separation-based implementation. For each monitored machine type, we select its normal recordings as targets and sample interference uniformly from all remaining normal recordings. We train a separator for each type to estimate the target recording from a mixture. With the separator frozen, we train the aggregation network on paired mixtures and their shared target recording. Differences between the paired embeddings provide the covariance for correction. At inference, the detector compares corrected embeddings with normal references for the monitored machine type. Section~\ref{sec:frozen_encoders} describes how the same aggregation objective and correction are applied to other frozen encoders.

For a fixed target machine type $m$, whose index is omitted except on the reference sets, let $\boldsymbol x$ denote a recording waveform and $j=1,\ldots,J$ index the acoustic units encoded from that recording. For the separator, each unit is a segment and the separator supplies $L$ intermediate feature maps per segment. The aggregation network combines them into an embedding $\boldsymbol z\in\mathbb R^{D_z}$. A training-only projection $\mathcal P_\psi$, with parameters $\psi$, maps $\boldsymbol z$ to $\boldsymbol r\in\mathbb R^{D_r}$ for representation learning. Anomaly scoring uses $\boldsymbol z$ before this projection.

\subsection{Separation features and regional aggregation}
\label{sec:separation_aggregation}

The separator follows the generator of the Conformer-based metric generative adversarial network (CMGAN)~\cite{cao2022cmgan}. Each training input combines a 2-s target segment with a 2-s non-target segment. Their root-mean-square levels determine the scaling for the selected SNR, and a common peak-scaling factor is applied to the mixture and target reference. The input is an uncompressed complex short-time Fourier transform (STFT), with its magnitude, real part, and imaginary part stacked as channels. A dense encoder and time--frequency Conformer blocks process the features. A magnitude-mask decoder and a complex-correction decoder jointly produce the estimated target spectrogram, from which the waveform is reconstructed.

Separation training minimizes
\begin{equation}
\begin{split}
\mathcal L_{\mathrm{sep}}
={}&\mathcal L_{\mathrm{logmag}}+\mathcal L_{\mathrm{powmag}}\\
&+0.05\mathcal L_{\mathrm{cplx}}+0.1\mathcal L_{\mathrm{negSI}}.
\end{split}
\label{eq:separation_loss}
\end{equation}
Here, $\mathcal L_{\mathrm{logmag}}$ is the mean absolute log-magnitude error, and $\mathcal L_{\mathrm{powmag}}$ is the MSE between magnitudes raised to the power of $0.3$. Both use the final estimated spectrogram and its target reference. The term $\mathcal L_{\mathrm{cplx}}$ averages Smooth L1 errors, with transition parameter $0.5$, over the real and imaginary components. The waveform term $\mathcal L_{\mathrm{negSI}}$ is the negative scale-invariant signal-to-distortion ratio~\cite{leroux2019sisdr}. Power compression is applied within the magnitude loss. The network input and output retain the uncompressed complex STFT.

\begin{figure*}[!t]
\centering
\includegraphics[width=0.95\linewidth]{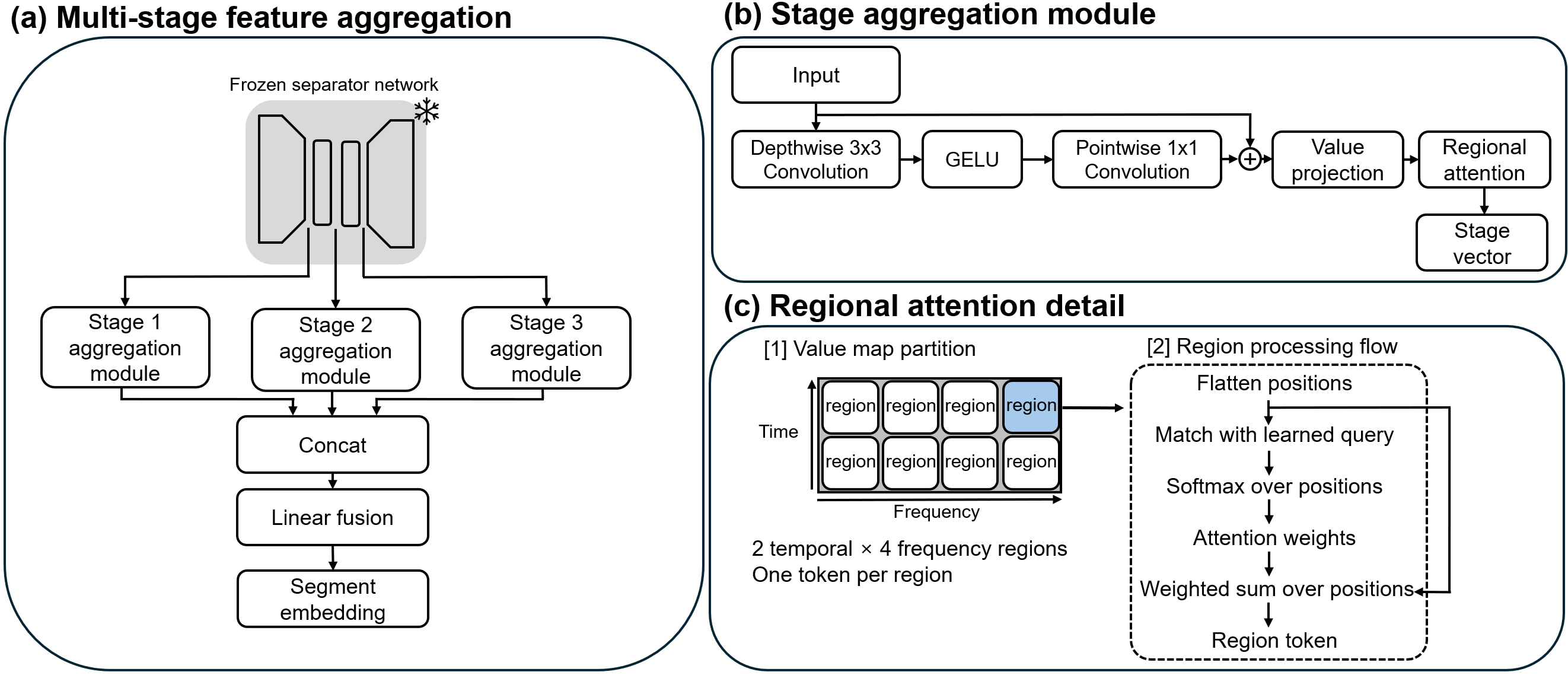}
\caption{Learnable aggregation of intermediate features in the separation-based implementation. (a) Stage-specific modules summarize the encoder and Conformer features, and linear fusion of their concatenated outputs produces a segment embedding. The example uses the encoder and two Conformer blocks. (b) Each module applies residual adaptation, value projection, and regional attention. (c) Learned-query attention pools positions within each region of a two-by-four time--frequency grid. The eight region vectors are concatenated and linearly mapped to a 128-dimensional stage vector before fusion across stages.}
\label{fig:regional_aggregation}
\end{figure*}

Fig.~\ref{fig:regional_aggregation} illustrates how features from the encoder and each Conformer block are aggregated. Let $\mathbf H^{(\ell)}\in\mathbb R^{C_\ell\times T_\ell\times F_\ell}$ denote the feature map at stage $\ell=1,\ldots,L$, where $C_\ell$, $T_\ell$, and $F_\ell$ are its channel, time, and frequency dimensions. Each stage is processed by a separate aggregation module. A residual adapter applies a $3\times3$ depthwise convolution, a Gaussian error linear unit (GELU), and a $1\times1$ pointwise convolution, then adds the result to $\mathbf H^{(\ell)}$. A further $1\times1$ value projection maps the channels to $D_{\mathrm{val}}=64$.

Regional pooling keeps coarse time--frequency locations separate, while attention learns the contributions of positions within each region. The projected map is divided into two temporal and four frequency intervals, forming $K=8$ non-overlapping regions that cover the map. For region $h=1,\ldots,K$, let $\mathbf V_{\ell,h}\in\mathbb R^{D_{\mathrm{val}}\times P_{\ell,h}}$ contain its $P_{\ell,h}$ position vectors. A learned query $\boldsymbol q_{\ell,h}\in\mathbb R^{D_{\mathrm{val}}}$ produces
\begin{equation}
\begin{aligned}
\boldsymbol w_{\ell,h}
&=\operatorname{softmax}
\left(\frac{\mathbf V_{\ell,h}^{\top}\boldsymbol q_{\ell,h}}
{\sqrt{D_{\mathrm{val}}}}\right),\\
\boldsymbol t_{\ell,h}
&=\mathbf V_{\ell,h}\boldsymbol w_{\ell,h}.
\end{aligned}
\label{eq:regional_attention}
\end{equation}
Here, $\boldsymbol w_{\ell,h}\in\mathbb R^{P_{\ell,h}}$ contains the position weights and $\boldsymbol t_{\ell,h}\in\mathbb R^{D_{\mathrm{val}}}$ is the region vector. The softmax operates over positions within the region, following scaled dot-product attention~\cite{vaswani2017attention}. Queries are shared across input segments, while the weights depend on the input features. Region vectors are concatenated in a fixed order and mapped by a stage-specific linear layer to a 128-dimensional stage vector. The $L$ stage vectors are concatenated and linearly mapped to $\boldsymbol z$, with $D_z=128$. The adapters, value projections, queries, and linear layers are trained jointly with the separator fixed.

\subsection{Paired-mixture representation learning}
\label{sec:paired_learning}

\begin{figure*}[!t]
\centering
\includegraphics[width=0.95\linewidth]{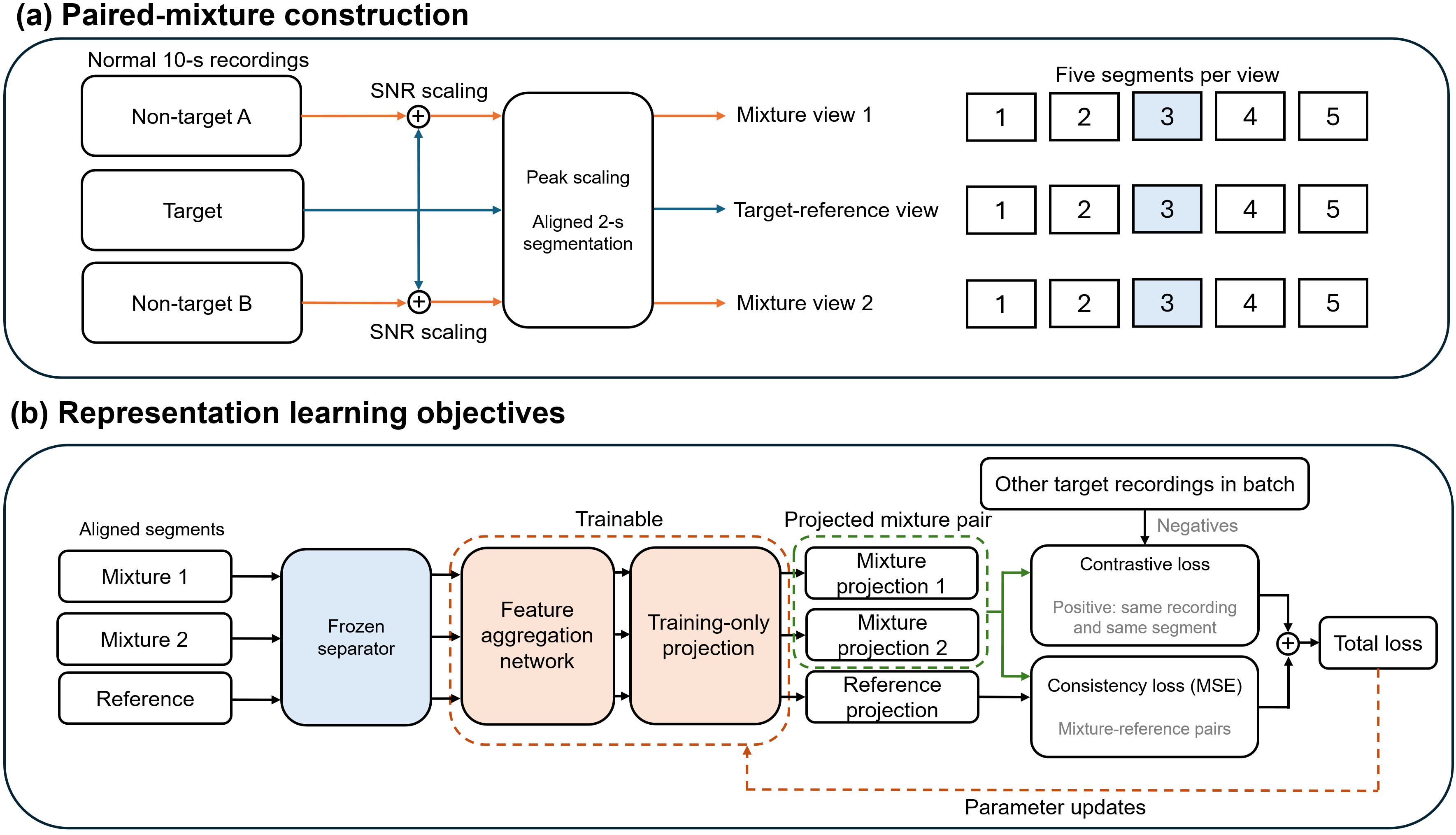}
\caption{Paired-mixture construction and representation learning for the separation-based implementation. (a) Two mixtures share a normal target recording and contain different non-target recordings. A common amplitude scale is applied to both mixtures and the target-reference view before division into five aligned 2-s segments. The highlighted segments indicate a corresponding triplet. (b) The frozen separator supplies features to the trainable aggregation network and training-only projection. Contrastive learning pairs mixture projections from the same target segment and uses mixture-segment projections from other target recordings in the batch as negatives. Consistency learning aligns each mixture projection with its target-reference projection.}
\label{fig:paired_mixture_learning}
\end{figure*}

Fig.~\ref{fig:paired_mixture_learning} illustrates the paired-mixture construction and representation learning objectives. Two mixture views share a normal target recording and use two distinct recordings from the non-target pool. Let $\boldsymbol u_i$ be target recording $i$ and $\boldsymbol n_i^{(v)}$ its sampled non-target recording for view $v\in\{1,2\}$. The views are
\begin{equation}
\begin{aligned}
\boldsymbol x_i^{(v)}
&=c_i\bigl(\boldsymbol u_i+\alpha_i^{(v)}\boldsymbol n_i^{(v)}\bigr),
\quad v=1,2,\\
\boldsymbol x_i^{(0)}&=c_i\boldsymbol u_i.
\end{aligned}
\label{eq:paired_views}
\end{equation}
The coefficient $\alpha_i^{(v)}$ sets the independently sampled SNR for view $v$, and $c_i$ is a common amplitude scale for both mixtures and the reference view. The original target recording retains its existing environmental sounds. Mixing is performed on 10-s recordings. For the separation-based implementation, all three views are divided at matching boundaries into $J=5$ non-overlapping 2-s segments.

A batch contains $B$ distinct target recordings, indexed by $i=1,\ldots,B$. For segment $j$ of view $\boldsymbol x_i^{(v)}$, the shared networks produce the embedding $\boldsymbol z_{i,j}^{(v)}$ and projection $\boldsymbol r_{i,j}^{(v)}=\mathcal P_\psi(\boldsymbol z_{i,j}^{(v)})$, with $v\in\{0,1,2\}$. The projection contains two linear layers with GELU between them, a hidden dimension of 256, and $D_r=128$ output dimensions.

For contrastive learning, $\mathcal Q$ contains the $2BJ$ mixture-segment indices $q=(i,j,v)$ with $v\in\{1,2\}$. The positive counterpart of $q$ is $p(q)=(i,j,3-v)$. The comparison set $\mathcal C(q)$ contains $p(q)$ and all segments from both mixture views of the other recordings in the batch. This assigns negatives at the recording level and excludes the remaining segments from recording $i$ and all target-reference projections. With $\operatorname{sim}$ denoting cosine similarity and temperature $\tau=0.2$, the loss is
\begin{equation}
\mathcal L_{\mathrm{ctr}}
=-\frac{1}{2BJ}\sum_{q\in\mathcal Q}
\log
\frac{\exp\!\left(\operatorname{sim}(\boldsymbol r_q,\boldsymbol r_{p(q)})/\tau\right)}
{\displaystyle\sum_{q'\in\mathcal C(q)}
\exp\!\left(\operatorname{sim}(\boldsymbol r_q,\boldsymbol r_{q'})/\tau\right)}.
\label{eq:contrastive_loss}
\end{equation}
This objective uses the normalized temperature-scaled contrastive form~\cite{chen2020simclr}.

The consistency objective aligns each mixture projection with its corresponding target-reference projection through a mean squared error,
\begin{equation}
\mathcal L_{\mathrm{cons}}
=\frac{1}{2BJD_r}
\sum_{i=1}^{B}\sum_{j=1}^{J}\sum_{v=1}^{2}
\left\|\boldsymbol r_{i,j}^{(v)}-\boldsymbol r_{i,j}^{(0)}\right\|_2^2.
\label{eq:consistency_loss}
\end{equation}
The loss is applied directly to the projection outputs and averaged over views, recordings, segments, and dimensions. Gradients flow through both mixture and reference paths to the shared aggregation network and projection. Their joint objective is
\begin{equation}
\mathcal L_{\mathrm{rep}}=\mathcal L_{\mathrm{ctr}}+\mathcal L_{\mathrm{cons}}.
\label{eq:representation_loss}
\end{equation}
The contrastive term pairs views of the same target segment and uses segments from other recordings of the same machine type as negatives. The consistency term aligns each mixture projection with its target-reference projection.

\subsection{Embedding correction and anomaly scoring}
\label{sec:correction_scoring}

Paired-mixture training constrains the projection outputs, while anomaly scoring uses pre-projection embeddings that can retain interference-related variation. Correction statistics are estimated separately for each machine type after network training. For $N_{\mathrm{cal}}$ normal target recordings, two mixture views are constructed as in~\eqref{eq:paired_views}. Differences between corresponding segment embeddings are
\begin{equation}
\boldsymbol\delta_{i,j}
=\boldsymbol z_{i,j}^{(1)}-\boldsymbol z_{i,j}^{(2)}.
\label{eq:paired_difference}
\end{equation}
The interference covariance is estimated as
\begin{equation}
\widehat{\boldsymbol\Sigma}_{\mathrm{nui}}
=\operatorname{Cov}_{\mathrm{LW}}
\left(\{\boldsymbol\delta_{i,j}\}_{i,j}\right),
\label{eq:interference_covariance}
\end{equation}
where $i=1,\ldots,N_{\mathrm{cal}}$, $j=1,\ldots,J$, and $\operatorname{Cov}_{\mathrm{LW}}$ denotes Ledoit--Wolf shrinkage estimation after centering the difference vectors by their own mean~\cite{ledoit2004wellconditioned}. The estimate $\widehat{\boldsymbol\Sigma}_{\mathrm{nui}}\in\mathbb R^{D_z\times D_z}$ describes embedding variation associated with changes in added non-target sounds for a shared target segment.

For machine type $m$, the reference set $\mathcal B_m=\{\boldsymbol z_n^{\mathrm{ref}}\}_{n=1}^{N_{\mathrm{bank}}}$ contains segment embeddings from all original normal training recordings of type $m$, pooled across sections and domains. Its mean defines the origin for the corrected cosine comparison. The mean and corrected embedding are given by
\begin{equation}
\begin{aligned}
\boldsymbol\mu
&=\frac{1}{N_{\mathrm{bank}}}\sum_{n=1}^{N_{\mathrm{bank}}}
\boldsymbol z_n^{\mathrm{ref}},\\
\widetilde{\boldsymbol z}
&=\widehat{\boldsymbol\Sigma}_{\mathrm{nui}}^{-1/2}
(\boldsymbol z-\boldsymbol\mu).
\end{aligned}
\label{eq:embedding_correction}
\end{equation}
Here, $\widehat{\boldsymbol\Sigma}_{\mathrm{nui}}^{-1/2}$ denotes the symmetric inverse square root~\cite{kessy2018optimalwhitening}. The same transformation is applied to inputs and references, yielding the corrected bank $\widetilde{\mathcal B}_m$.

At inference, the separation-based implementation processes the $J$ segments of the input recording using the fixed separator and aggregation network. For a corrected segment embedding $\widetilde{\boldsymbol z}$, cosine 1-nearest-neighbor (1-NN) scoring assigns
\begin{equation}
s(\widetilde{\boldsymbol z};\widetilde{\mathcal B}_m)
=\min_{\widetilde{\boldsymbol b}\in\widetilde{\mathcal B}_m}
\left(1-\frac{\widetilde{\boldsymbol z}^{\top}\widetilde{\boldsymbol b}}
{\|\widetilde{\boldsymbol z}\|_2\|\widetilde{\boldsymbol b}\|_2}\right).
\label{eq:segment_score}
\end{equation}
The recording-level score is
\begin{equation}
\mathcal A(\boldsymbol x)
=\frac{1}{J}\sum_{j=1}^{J}
s(\widetilde{\boldsymbol z}_j;\widetilde{\mathcal B}_m).
\label{eq:recording_score}
\end{equation}
Larger scores indicate greater average distance from the nearest normal references.

\subsection{Regional aggregation with other frozen encoders}
\label{sec:frozen_encoders}

The same paired-mixture objective can train a regional readout for features from other encoders. For BEATs, EAT, and CED, we take the final encoder patch features in their native time--frequency arrangement. For ResNets trained through classification or self-supervision, we use the final convolutional map before global pooling. Each map is treated as one stage ($L=1$) and divided into two temporal and four frequency regions as in Section~\ref{sec:separation_aggregation}. For the reconstruction autoencoder, we extract the sequence of fourth-block encoder activations before the bottleneck by passing overlapping five-frame windows through the network. The sequence retains temporal order, so we divide it into eight temporal regions. Each regional readout produces a 128-dimensional embedding.

The original input preprocessing, encoder weights, and normalization statistics remain fixed. For each target machine type, the aggregation network and training-only projection learn from the two 10-s mixtures and their shared target-reference recording. Here, the full recording is the unit of comparison ($J=1$) in \eqref{eq:contrastive_loss} and~\eqref{eq:consistency_loss}. Each paired recording supplies one embedding difference for covariance estimation and each original normal training recording supplies one reference. Equation~\eqref{eq:recording_score} therefore gives a single cosine nearest-reference distance per recording. We compare the learned readout with mean pooling over the same frozen features. Regional aggregation with and without correction uses the same trained aggregation checkpoint.

\section{Experimental setup}
\label{sec:setup}

\subsection{Data and evaluation protocol}
\label{sec:data_protocol}

Experiments use DCASE 2022 Task 2 recordings from seven machine types (bearing, fan, gearbox, slider, ToyCar, ToyTrain, and valve)~\cite{dohi2022task2overview,dohi2022mimiidg,harada2021toyadmos2}. Each recording is a 10-s, single-channel waveform sampled at 16~kHz. Within each machine type, a section groups recordings for a domain-shift scenario and defines a unit of performance evaluation. The source domain represents the conditions accounting for most normal training recordings, while the target domain covers different conditions represented by a small normal training subset. These shifts arise from changes in operating or recording conditions, such as speed, load, microphone arrangement, or environmental noise. Both domains contain normal and anomalous test recordings~\cite{dcase2022task2}.

Normal training recordings combine the development and additional training sets~\cite{dcase2022devdata,dcase2022traindata}, covering sections 00--05. The specified allocation is 990 source-domain and 10 target-domain normal recordings per section. The available pool contains 5,999 recordings for bearing and 6,000 for each other type. Within each machine type, we pool normal recordings across all six sections and both domains. This pool supplies the target recordings for training, the calibration subset, and the normal references for scoring. In mixture construction, the target component is a recording of the monitored machine type, whose pool includes both source and target domains. Non-target components come from other machine types.

Development-test sections 00--02 provide a common evaluation set for the detector and scorer comparisons. The interference tests use mixtures constructed from the same test recordings. Each section contains 100 normal and 100 anomalous recordings across the two domains, giving 600 test recordings per machine type and 4,200 in total. Each test recording is scored by the detector and normal references for the monitored target. Section and domain labels, together with test normal/anomalous labels, define the evaluation subsets.

\subsection{Implementation and training}

Separation inputs use a 400-sample Hamming window, a 400-point fast Fourier transform (FFT), and a 200-sample hop. The target-to-non-target SNR is sampled uniformly from $-5$ to $5$~dB in 1~dB steps. Separator training constructs mixtures from randomly selected 2-s target and non-target segments. For CMGAN, aggregation training constructs the three 10-s views in~\eqref{eq:paired_views} and divides them into five aligned 2-s segments.

Both stages use AdamW~\cite{loshchilov2019decoupled}. The separator is trained for 200 epochs at an initial learning rate of $5\times10^{-4}$, halved every 40 epochs, with a batch size of 8 (6 for the four-block, 128-channel configuration). Each time- and frequency-axis Conformer uses four attention heads, a feed-forward expansion factor of four, and a depthwise-convolution kernel length of 31. Attention and feed-forward dropout are 0.2, and convolution-module dropout is 0.

The aggregation network and projection are trained for 20 epochs with a constant learning rate of $10^{-3}$ and $B=16$ distinct target recordings per batch. For CMGAN, each view supplies 80 segment projections, and the two mixture views contribute 160 projections to the contrastive objective. In both stages, one epoch is one pass through the corresponding machine type's normal target-recording list. Checkpoints are selected by minimum epoch-averaged total training loss. Task-specific comparison checkpoints follow the same criterion. Dis-multi-PL uses the final checkpoint of its reference recipe.

Separator configurations vary the Conformer block count from zero to four and the channel count between 64 and 128. The zero-block configuration retains the encoder and decoders, with the encoder output as the sole extracted feature stage. The representative CMGAN model has two blocks and 64 channels, with 128-dimensional segment embeddings. For covariance estimation, 512 normal target recordings per type are selected at evenly spaced indices in the training list. For each selected recording, two mixture views are generated by independently sampling the non-target recording and SNR. Five aligned segments yield 2,560 segment-difference vectors per machine type for CMGAN. The reference mean and bank use all original normal training recordings.

We repeat training and evaluation for CMGAN, direct paired-mixture ResNet-18, and Dis-multi-PL using three seeds per system. For BEATs iter3, three runs vary the seed for aggregation training (1, 2, and 3) with the same pre-trained encoder. For these runs, the mean and sample standard deviation of overall harmonic mean (\(H\)) are reported.

\subsection{Comparison systems}
\label{sec:comparisons}

For the metadata-free comparison, we selected methods that can operate with normal recordings grouped by machine type. The proposed method uses this grouping to select target recordings. The comparison methods can also use machine type as a classification target or derive pseudo-labels from the audio. We report the use of externally pre-trained models separately in Table~\ref{tab:overall_representative}.

We train a reconstruction autoencoder following the DCASE baseline~\cite{dcase2022task2}. For the ResNet backbones~\cite{he2016resnet}, we use cross-entropy (CE), ArcFace~\cite{deng2019arcface}, SimCLR~\cite{chen2020simclr}, or SimSiam~\cite{chen2021simsiam}. The autoencoder is trained separately for each machine type. Classification and self-supervised ResNets use recordings pooled across types, with machine types providing the classification labels and augmented views providing the self-supervised relationships. These systems use 128-bin log-mel inputs computed with a 1024-point FFT and a 512-sample hop. The autoencoder reconstructs five-frame groups, while the ResNets process 10-s recordings. Classification features are extracted before the final layer, and backbone and projection features are evaluated for SimCLR and SimSiam.

Direct paired-mixture learning applies the objectives of Section~\ref{sec:paired_learning} to jointly trained ResNet-18, -34, -50, -101, and -152 features. Each backbone is trained separately for each target type using 2-s inputs and 128-dimensional embeddings. The backbone, embedding layer, and projection are jointly trained for 200 epochs with AdamW, a learning rate of $10^{-3}$, and 16 target recordings per batch. Separation-feature comparisons use GAP, learned aggregation, and embedding correction with the same frozen separator checkpoint and extraction stages. GAP is a parameter-free readout that averages each channel over time and frequency and concatenates the stage vectors. Learned aggregation produces 128-dimensional embeddings.

The external-model comparisons use frozen BEATs~\cite{chen2023beats}, EAT~\cite{chen2024eat}, and CED~\cite{dinkel2024ced} features with their original preprocessing. For BEATs and EAT, we follow the pooling strategy used by Wilkinghoff et al.~\cite{wilkinghoff2026local} and evaluate fixed readouts that average patch features over time and retain frequency position. For EAT, we evaluate the classification-token (CLS) feature alone and concatenated with these features. For Dis-multi-PL~\cite{fujimura2026pseudolabel}, we cluster BEATs iter3 patch means by machine type, obtaining 335 pseudo-label classes on our training data. These labels supervise a single student across machine types. At evaluation, the 512-dimensional student embedding is used for normal-reference construction and scoring. The supplementary material provides configuration details, machine-wise comparisons, and computational cost estimates.

\subsection{Matched readouts for frozen encoders}
\label{sec:frozen_encoder_protocol}

To test how aggregation and correction depend on the input representation, we evaluate eight pre-trained audio encoder configurations (BEATs, EAT, and CED), 20 ResNet configurations (five depths and four training objectives), and the autoencoder with hidden and latent dimensions of 128 and 8. The fixed-readout entries in Table~\ref{tab:overall_representative} use the best-performing feature construction within each encoder family.

For each machine type, we train the aggregation network and projection using the same 20-epoch schedule, mixture SNR range, and batch size as for CMGAN aggregation. AdamW uses weight decay 0.01. We select the checkpoint with the lowest mean training loss per epoch and evaluate it before and after correction. The matched readout comparisons use aggregation seed 1.

With one embedding per recording, the 512 selected normal targets yield 512 difference vectors per machine type for covariance estimation. Centering and cosine 1-NN use the full set of original normal references.

\subsection{Metrics and comparison protocol}
\label{sec:metrics}

For the overall and machine-wise ASD comparisons, embedding-based systems use cosine 1-NN with the normal reference bank of the corresponding machine type, while the reconstruction autoencoder uses reconstruction MSE. The autoencoder readout comparison uses cosine 1-NN on its recording embeddings. Recording-level embeddings are scored directly, and the CMGAN and direct paired-mixture systems average five segment scores.

Detection performance is evaluated using the area under the receiver operating characteristic curve (AUC) and standardized partial AUC (pAUC). Source and target AUCs compare normal recordings from the corresponding domain with all anomalous recordings from both domains in the same section. Following the DCASE protocol~\cite{dohi2022task2overview}, we use all normal and anomalous recordings and a maximum false positive rate of 0.1 for the standardized pAUC~\cite{mcclish1989portion}. Each metric is summarized by the harmonic mean over the three sections of each machine type. The overall value for each metric is the harmonic mean over all 21 sections. The machine-wise harmonic mean $H$ combines the nine metric values from three sections. The overall $H$ combines all 63 values across seven machine types. Results are expressed as percentages, with differences calculated before rounding.

Table~\ref{tab:overall_representative} reports the highest development-test $H$ among the configurations compared for each method. The CMGAN comparison varies separator depth and width and uses a $2\times4$ grid for regional aggregation. For each method, the selected configuration is shared across all seven machine types. Because configuration selection and evaluation use the same development-test recordings, the resulting scores may be optimistic. Table~\ref{tab:seed_variability} reports variation across three training runs of four selected systems on these same recordings.

\subsection{Controlled acoustic interference at test time}
\label{sec:interference_protocol}

Each of the 4,200 test recordings is mixed with a normal test recording from another machine type or with Gaussian white noise. Mixing recordings simulates overlap between the operating sounds of two machines. White noise provides a broadband comparison. For each 10-s input, the added waveform is scaled to give an original-recording-to-interference power ratio of $-9$, $-6$, $-3$, $0$, $+3$, $+6$, or $+9$~dB. The unmodified recording retains its existing background sound. Its normal or anomalous label applies to every mixture. For a given input and interference type, all systems receive the same added waveform at each level, and only its amplitude changes across SNRs.

We evaluate the representative checkpoints with their original normal reference banks, correction statistics, and cosine 1-NN scorers held fixed. The overall $H$ is computed separately for each interference type and SNR.

\section{Results and Discussion}
\label{sec:results}
\label{sec:analysis}

\subsection{Overall detection performance}
\label{sec:overall_results}

Table~\ref{tab:overall_representative} compares representative systems on the original development-test recordings. For each method, one configuration is shared across all seven machine types. External pre-training supplies either frozen features for scoring or a teacher for pseudo-label generation.

\begin{table*}[t]
\centering
\caption{Overall $H$ on the original development-test recordings under the metadata-free protocol. Each row gives the highest development-test $H$ among the configurations we evaluated for that method in the original comparison run. CMGAN depth and width are selected separately for each readout. Regional aggregation uses a $2\times4$ grid.}
\label{tab:overall_representative}
\small
\begin{tabular*}{\textwidth}{@{\extracolsep{\fill}}llr@{}}
\toprule
Method & Representative configuration & $H$ (\%) \\
\midrule
\multicolumn{3}{@{}l}{\textit{Without externally pre-trained models}} \\
Autoencoder~\cite{dcase2022task2} & Hidden 256, latent 16 (reconstruction MSE) & 57.20 \\
CE & ResNet-50~\cite{he2016resnet}, backbone & 60.74 \\
ArcFace~\cite{deng2019arcface} & ResNet-152, backbone & 58.08 \\
SimCLR~\cite{chen2020simclr} & ResNet-34, projected features & 55.34 \\
SimSiam~\cite{chen2021simsiam} & ResNet-152, backbone & 54.56 \\
Direct paired-mixture learning & ResNet-18, before correction & 63.59 \\
Separation + GAP & 3 blocks, 128 channels & 60.13 \\
Separation + learned aggregation & 2 blocks, 128 channels & 63.12 \\
CMGAN + aggregation + correction & 2 blocks, 64 channels & \textbf{64.20} \\
\midrule
\multicolumn{3}{@{}l}{\textit{With externally pre-trained models}} \\
BEATs, fixed readout~\cite{chen2023beats} & iter3, time mean + frequency concatenation & 62.61 \\
EAT, fixed readout~\cite{chen2024eat} & Large, CLS + time mean + frequency concatenation & 60.33 \\
CED, fixed readout~\cite{dinkel2024ced} & Small, patch mean & 57.80 \\
Dis-multi-PL~\cite{fujimura2026pseudolabel} & Student trained with BEATs pseudo-labels & 65.03 \\
BEATs + aggregation + correction & iter3 & \textbf{65.35} \\
\bottomrule
\end{tabular*}
\end{table*}

Among the systems trained without external pre-training, the two-block, 64-channel CMGAN model achieved the highest $H$ (64.20\%) on the original recordings. Direct paired-mixture learning reached 63.59\%. Both achieve higher $H$ than the reconstruction, machine-type classification, SimCLR, and SimSiam systems evaluated here. Machine-type classification learns distinctions between types, while paired mixtures provide correspondences between views of the same target segment under changes in added sound. The performance of both paired-mixture systems supports using these recording-derived relationships to learn features for within-type anomaly detection.

\begin{table}[t]
\centering
\caption{Overall $H$ across three training runs on the same development-test set, including the original comparison run. SD is the sample standard deviation in percentage points.}
\label{tab:seed_variability}
\footnotesize
\setlength{\tabcolsep}{2pt}
\begin{tabular*}{\columnwidth}{@{\extracolsep{\fill}}>{\raggedright\arraybackslash}p{.53\columnwidth}rr@{}}
\toprule
System & Mean $H$ (\%) & SD (pp) \\
\midrule
Direct paired-mixture learning (ResNet-18) & 62.33 & 1.45 \\
CMGAN + aggregation + correction & 64.56 & 0.33 \\
Dis-multi-PL & 64.87 & 0.87 \\
BEATs iter3 + aggregation + correction & 64.88 & 0.62 \\
\bottomrule
\end{tabular*}
\end{table}

Across three runs, CMGAN achieved a mean $H$ of 64.56\%, compared with 62.33\% for direct paired-mixture learning and 64.87\% for Dis-multi-PL (Table~\ref{tab:seed_variability}). Regional aggregation with embedding correction on frozen BEATs iter3 features achieved 65.35\% in the representative run, compared with 65.03\% for Dis-multi-PL using pseudo-labels derived from the same encoder (Table~\ref{tab:overall_representative}). The corresponding three-run means were 64.88\% and 64.87\%.

\subsection{Contributions of aggregation and correction}
\label{sec:aggregation_results}

With the two-block, 64-channel separator checkpoint, extraction stages, normal reference recordings, and cosine 1-NN scorer fixed, $H$ increases from 58.15\% with GAP to 61.85\% with learned aggregation and to 64.20\% after correction. The 3.70-percentage-point aggregation gain reflects the full learned readout, whose local adaptation, regional pooling, and stage fusion are jointly trained through paired-mixture correspondence and target-reference consistency. Subsequent correction adds 2.35 points. Across the ten separator configurations, learned aggregation exceeds GAP in nine and corrected aggregation exceeds GAP in all ten. For the four-block, 64-channel separator, correction reduces $H$ from 62.81\% to 61.45\%. Supplementary Table~S2 reports all ten configurations.

\begin{table}[!t]
\centering
\caption{Effect of the correction components for the same two-block, 64-channel CMGAN separator. Both embeddings use cosine 1-NN. Values are overall $H$ (\%) on the original development-test recordings.}
\label{tab:correction_components}
\footnotesize
\setlength{\tabcolsep}{2pt}
\begin{tabular*}{\columnwidth}{@{\extracolsep{\fill}}>{\raggedright\arraybackslash}p{.56\columnwidth}rr@{}}
\toprule
Transformation & GAP & \shortstack[r]{Learned\\aggregation} \\
\midrule
None & 58.15 & 61.85 \\
Mean centering & 58.61 & 62.67 \\
Interference covariance & 59.52 & 62.38 \\
Normal covariance + centering & 60.08 & 60.30 \\
Interference covariance + centering & 59.88 & 64.20 \\
\bottomrule
\end{tabular*}
\end{table}

For the learned embeddings, centering and interference covariance each change $H$ by less than one percentage point when applied alone (Table~\ref{tab:correction_components}). With centering in both conditions, interference covariance yields 64.20\% and normal covariance yields 60.30\%. Both estimates use 512 normal target recordings. Normal covariance summarizes variation across original target recordings, whereas paired-mixture covariance describes variation associated with changing the added sound for a shared target segment. For GAP features, centering with either covariance estimate yields similar performance.

Across three configurations selected using a preliminary cosine 5-NN evaluation, target-component estimation produces larger correction gains than non-target estimation when evaluated with cosine 1-NN (Supplementary Table~S3). For the two-block, 64-channel separator, the corresponding changes are from 61.85\% to 64.20\% and from 61.15\% to 62.15\%, respectively. Target estimation uses the shared component as the separation reference, and the subsequent paired views preserve this component while changing the added sound.

Correction increased CMGAN $H$ for six of the seven machine types (Supplementary Table~S1). Bearing showed the largest gain, from 60.31\% to 68.60\%, with improvements in source AUC, target AUC, and pAUC. For slider, correction reduced $H$ from 76.38\% to 72.39\%, with the largest decrease in target AUC (72.00\% to 66.16\%). For ToyTrain, target AUC increased from 31.72\% to 32.98\% after correction but remained well below the source AUC of 84.62\%. Supplementary Table~S18 gives the corresponding machine- and domain-wise results for BEATs.

\subsection{Anomaly scoring and recording-level score distributions}
\label{sec:scorers}

Table~\ref{tab:score_sensitivity} compares five scorers using the same two-block, 64-channel CMGAN checkpoint and normal reference recordings. We score GAP and regional embeddings before and after correction. The Mahalanobis distance (MD) scorers use the mean of the normal references and either their empirical covariance (MD-Emp) or a Ledoit--Wolf shrinkage estimate (MD-LW)~\cite{ledoit2004wellconditioned}. Embedding correction uses a separate covariance estimated from paired-mixture differences. We compare these scores with cosine 1-NN, cosine 5-NN, and local density normalization~\cite{wilkinghoff2026local}. The local scale uses five neighbors from other training recordings (Supplementary Table~S15 and accompanying text).

\begin{table}[!t]
\caption{Overall $H$ (\%) on the original recordings in sections 00--02 for the same CMGAN checkpoint. Corr. denotes embedding correction.}
\label{tab:score_sensitivity}
\centering
\footnotesize
\begin{tabular*}{\columnwidth}{@{\extracolsep{\fill}}lrrrr@{}}
\toprule
& \multicolumn{2}{c}{GAP} & \multicolumn{2}{c}{Regional} \\
Scorer & Raw & Corr. & Raw & Corr. \\
\midrule
MD-Emp & 54.13 & 54.13 & 51.30 & 51.30 \\
MD-LW & 54.29 & 54.18 & 49.67 & 51.14 \\
Cosine 1-NN & 58.15 & 59.88 & 61.85 & 64.20 \\
Cosine 5-NN & 56.57 & 58.25 & 60.51 & 62.82 \\
Local density & 59.93 & 58.59 & 59.55 & 62.42 \\
\bottomrule
\end{tabular*}
\end{table}

For corrected regional embeddings, normalizing the input and reference embeddings to unit length and refitting the normal mean and shrinkage covariance raised MD-LW from 51.14\% to 57.11\%. Cosine 1-NN achieved 64.20\% on the same unit vectors, 7.09 percentage points above the refitted MD-LW score.

\begin{figure*}[!t]
\centering
\includegraphics[width=0.95\textwidth]{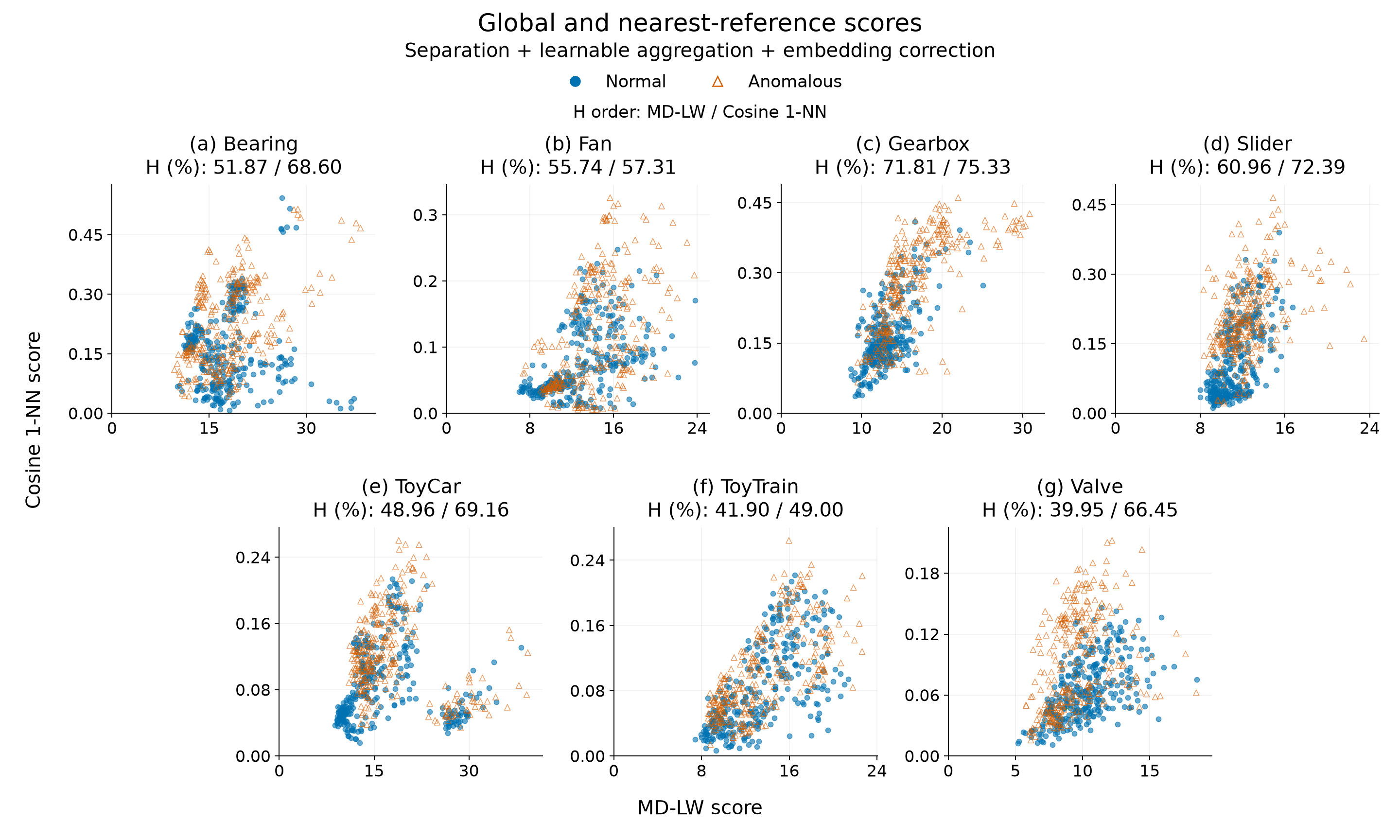}
\caption{Mahalanobis and cosine 1-NN scores on the original recordings in sections 00--02 for the corrected regional CMGAN embeddings. Each panel shows 300 normal and 300 anomalous recordings of one machine type. Each point is a 10-s recording scored by averaging five segment scores. The $H$ values in each panel are for MD-LW and cosine 1-NN, respectively. Axis limits are set separately for each machine type.}
\label{fig:score_distributions_c64}
\end{figure*}

For each machine type, MD-LW uses a single mean and covariance for normal references pooled across sections and domains. Some normal bearing and ToyCar recordings have large MD-LW scores yet match individual references closely in embedding direction (Fig.~\ref{fig:score_distributions_c64}). For valve, anomalous recordings with MD-LW scores in the normal range tend to receive larger cosine 1-NN scores. For this machine, $H$ was 39.95\% for MD-LW and 66.45\% for cosine 1-NN. Fan and ToyTrain show substantial overlap under both scorers.

\subsection{Regional aggregation on other frozen encoders}
\label{sec:encoder_results}

Table~\ref{tab:pretrained_regional_detection} compares mean pooling and regional aggregation on the same frozen encoder features. The corrected and uncorrected regional embeddings use the same aggregation checkpoint.

\begin{table}[t]
\centering
\caption{Overall $H$ (\%) for eight externally pre-trained encoders with fixed weights. Mean averages the final patch features. Regional learns a readout of the same features. Correction is applied to its output. All columns use cosine 1-NN.}
\label{tab:pretrained_regional_detection}
\footnotesize
\setlength{\tabcolsep}{2pt}
\begin{tabular*}{\columnwidth}{@{\extracolsep{\fill}}lrrr@{}}
\toprule
Frozen encoder & Mean & Regional & \shortstack[r]{Regional\\+ correction} \\
\midrule
CED-tiny & 56.88 & 60.25 & 61.23 \\
CED-mini & 57.38 & 61.97 & 62.54 \\
CED-small & 57.80 & 62.39 & 62.49 \\
CED-base & 57.68 & 63.70 & 61.36 \\
EAT-base & 58.07 & 60.22 & 61.44 \\
EAT-large & 59.68 & 60.01 & 61.81 \\
BEATs iter3 & 60.81 & 63.76 & 65.35 \\
BEATs iter3+ (AS2M) & 60.26 & 61.85 & 64.50 \\
\bottomrule
\end{tabular*}
\end{table}

Regional aggregation improves $H$ over mean pooling for all eight pre-trained encoders. For BEATs iter3, $H$ increases from 60.81\% to 63.76\% with regional aggregation and to 65.35\% after correction, compared with 62.61\% for the fixed readout that retains frequency position. Correction further improves $H$ for seven encoders. For CED-base, $H$ falls from 63.70\% to 61.36\%. The mean and learned readouts differ in output dimension.

At all five SimCLR backbone depths, regional aggregation gives the highest $H$, and the corrected embeddings still outperform mean pooling. The covariance transform in~\eqref{eq:embedding_correction} gives less weight to directions in which the differences between paired embeddings have higher variance. For some encoders, these directions may also help distinguish normal from anomalous recordings. Across the twenty ResNet configurations, regional aggregation with correction exceeds mean pooling in eleven. Supplementary Table~S16 also reports the autoencoder results.

\subsection{Detection under added acoustic interference}
\label{sec:noise_results}

\begin{figure*}[!t]
\centering
\includegraphics[width=0.65\textwidth]{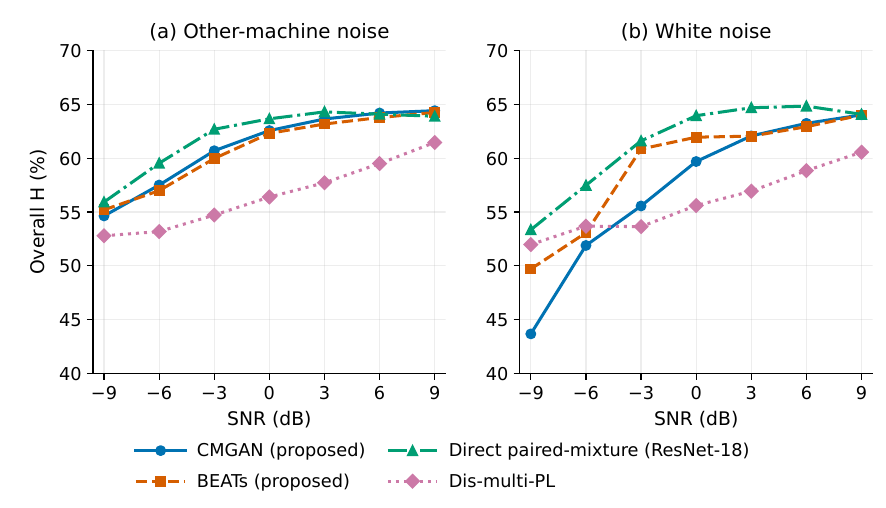}
\caption{Overall harmonic mean $H$ for the four representative checkpoints on development-test sections 00--02 with (a) a normal recording from another machine type and (b) Gaussian white noise. The SNR gives the power of the original recording relative to the added sound. Both panels use the same vertical scale.}
\label{fig:noise_snr_response}
\end{figure*}

Fig.~\ref{fig:noise_snr_response} compares the CMGAN and BEATs implementations with direct paired-mixture ResNet-18 and Dis-multi-PL across seven SNRs using cosine 1-NN (Section~\ref{sec:interference_protocol}). Their respective $H$ values on the original recordings were 64.20\%, 65.35\%, 63.59\%, and 65.03\%. Each curve uses one fixed checkpoint.

With other-machine interference, both proposed implementations exceeded Dis-multi-PL at all seven SNRs. Among the four systems in Fig.~\ref{fig:noise_snr_response}, direct paired-mixture ResNet-18 gave the highest $H$ from $-9$ to $+3$~dB, and CMGAN gave the highest $H$ at $+6$ and $+9$~dB. At 0~dB, $H$ was 62.58\% for CMGAN, 62.32\% for BEATs, 63.67\% for direct ResNet-18, and 56.40\% for Dis-multi-PL. Relative to the original recordings, $H$ fell by 1.62 percentage points for CMGAN, 3.03 for BEATs, and 8.63 for Dis-multi-PL.

With Gaussian white noise, direct ResNet-18 led the four systems at all seven SNRs. Both proposed implementations exceeded Dis-multi-PL from $-3$ to $+9$~dB. Dis-multi-PL scored higher than both at $-9$ and $-6$~dB.

Direct ResNet-18 and the CMGAN readout share the paired-mixture objective. On the original recordings, correction raised $H$ for the regional CMGAN embeddings from 61.85\% to 64.20\%, above the 63.59\% obtained by direct ResNet-18 in the same comparison. The covariance used for correction was estimated from mixtures of normal machine recordings. At $-9$~dB, applying the same correction to the fixed regional embeddings raised $H$ from 52.59\% to 54.64\% with other-machine interference and lowered it from 48.91\% to 43.66\% with white noise (Supplementary Table~S19). Supplementary Tables~S20 and~S21 give the results for the additional encoders.

\section{Conclusion}
\label{sec:conclusion}

Paired mixtures sharing a normal target recording provided supervision for regional aggregation and a covariance estimate for embedding correction. For the regional embeddings of the fixed separator, correction using paired-mixture covariance gave a higher $H$ than correction using normal covariance when both used the same normal-reference mean. On the original development-test recordings, CMGAN ranked first among the systems evaluated without external pre-training. Its three-run mean $H$ of 64.56\% was close to that of Dis-multi-PL and 2.23 percentage points above direct paired-mixture learning. With BEATs iter3, regional aggregation with embedding correction achieved a higher $H$ in the representative run than Dis-multi-PL, which used the same encoder to derive pseudo-labels for its student. The two methods had similar three-run means.

With the representative checkpoints held fixed, both proposed systems achieved a higher $H$ than Dis-multi-PL at all seven levels of other-machine interference. Under Gaussian white noise, direct paired-mixture learning achieved the highest $H$ of the four systems at every tested level. These results support training with mixtures of normal machine recordings when interference from other machines is expected.

\section*{Acknowledgment}
The authors used ChatGPT (OpenAI) for language editing throughout the manuscript and assistance with code development.

\bibliographystyle{IEEEtran}
\bibliography{references}

\begin{IEEEbiography}[{\includegraphics[width=1in,height=1.25in,clip,keepaspectratio]{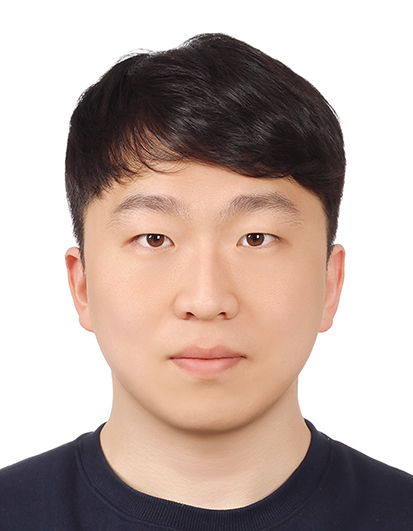}}]{Seunghyeon Shin}
received the B.S. degree in mechanical engineering from Dong-A University, Busan, Republic of Korea, in 2020, and the M.S. degree in electronic engineering from Kyungpook National University, Daegu, Republic of Korea, in 2022. He is currently pursuing the Ph.D. degree with the School of Electronic and Electrical Engineering, Kyungpook National University.

His research interests include anomalous sound detection, acoustic source separation, and acoustic signal processing.
\end{IEEEbiography}

\begin{IEEEbiography}[{\includegraphics[width=1in,height=1.25in,clip,keepaspectratio]{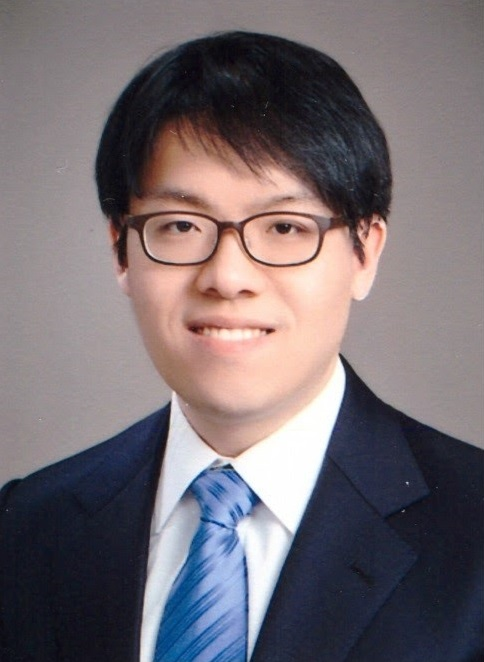}}]{Seokjin Lee}
(Member, IEEE) received the B.S., M.S., and Ph.D. degrees in electrical and computer engineering from Seoul National University, Seoul, Republic of Korea, in 2006, 2008, and 2012, respectively.

From 2012 to 2014, he was a Senior Research Engineer with LG Electronics. From 2014 to 2018, he was an Assistant Professor with the Department of Electronics Engineering, Kyonggi University, Suwon, Republic of Korea. He joined Kyungpook National University, Daegu, Republic of Korea, in 2018, where he is currently a Professor. His research interests include signal processing and deep learning for acoustic, audio, and music signals.
\end{IEEEbiography}
\EOD
\end{document}


\begin{center}
{\Large\bfseries Supplementary Material}\par\medskip
{\large Paired-Mixture Learning and Embedding Correction for Metadata-Free Anomalous Machine Sound Detection}\par\medskip
Seunghyeon Shin and Seokjin Lee
\end{center}

Detection results in this supplement use development-test sections 00--02 of DCASE 2022 Task 2. Tables~\ref{tab:supp_proposed_metrics}--\ref{tab:supp_pretrained} report the original-run results underlying the system and configuration comparisons. Table~\ref{tab:appendix_encoder_repeats} reports repeated training runs on the original development-test recordings. All detection metrics are expressed as percentages. Source-domain AUC, target-domain AUC, and pAUC are summarized by harmonic means over the three sections for each machine type and over all 21 machine--section combinations for the overall values. The machine-wise $H$ combines nine section-level metrics, and overall $H$ combines all 63 metrics. Differences are calculated before rounding. Src.\ and Tgt.\ denote source- and target-domain AUCs, respectively.

\section{Detailed Experimental Results}
\label{supp:results}

Table~\ref{tab:supp_proposed_metrics} reports GAP, learned aggregation, and the corrected representation from the representative two-block, 64-channel separator. Tables~\ref{tab:supp_separator_grid}--\ref{tab:supp_scorers} expand the architecture, separation-target, correction-component, and scorer comparisons for the task-trained CMGAN features. The configurations and scoring rules follow the definitions in the main text.

Tables~\ref{tab:supp_comparisons_a}--\ref{tab:supp_comparisons_d} give the machine-wise metrics for the comparison systems evaluated on the original development-test recordings. Tables~\ref{tab:supp_ae_direct}--\ref{tab:supp_pretrained} list the evaluated model and feature configurations. For these comparison systems, the autoencoder uses reconstruction MSE, and the embedding-based systems use cosine 1-NN. Backbone and projection features use the same checkpoint for each self-supervised model. Frequency concatenation retains the frequency positions after temporal averaging.

The separator grid varies the number of Conformer blocks (0--4) and the channel count (64 or 128). Each architecture uses the same frozen checkpoint for all readouts. The three architectures used to compare target-component and non-target-component estimation were selected by the highest corrected $H$ in a preliminary cosine 5-NN comparison of the target-estimation systems. Table~\ref{tab:supp_objectives} evaluates those architectures with cosine 1-NN. For the correction ablation, both covariance estimators use the same 512 normal target recordings and Ledoit--Wolf shrinkage. The normal-covariance estimate uses original embeddings, while the interference-covariance estimate uses differences between paired mixtures sharing the target segment. Input and reference embeddings receive the same transformation.

\begin{table}[!htbp]
\centering
\caption{Machine-wise metrics of the two-block, 64-channel pipeline}
\label{tab:supp_proposed_metrics}
\footnotesize
\setlength{\arrayrulewidth}{0.3pt}
\setlength{\tabcolsep}{2.4pt}
\renewcommand{\arraystretch}{1.12}
\begin{tabular*}{\textwidth}{@{\extracolsep{\fill}}lrrrrrrrrrrrr@{}}
\hline
Machine & \multicolumn{4}{c}{GAP} & \multicolumn{4}{c}{Learned aggregation} & \multicolumn{4}{c}{Regional + correction} \\
\cline{2-5} \cline{6-9} \cline{10-13}
 & Src. & Tgt. & pAUC & $H$ & Src. & Tgt. & pAUC & $H$ & Src. & Tgt. & pAUC & $H$ \\
\hline
Bearing & 63.47 & 57.15 & 56.33 & 58.82 & 57.06 & 67.74 & 57.29 & 60.31 & 65.85 & 78.92 & 62.98 & 68.60 \\
\hline
Fan & 78.51 & 39.94 & 54.61 & 53.49 & 79.54 & 42.64 & 56.65 & 55.89 & 79.76 & 44.05 & 58.44 & 57.31 \\
\hline
Gearbox & 83.59 & 75.93 & 66.53 & 74.69 & 75.55 & 75.67 & 64.51 & 71.51 & 78.73 & 79.49 & 68.76 & 75.33 \\
\hline
Slider & 87.12 & 60.30 & 58.90 & 66.61 & 97.14 & 72.00 & 66.24 & 76.38 & 96.16 & 66.16 & 62.77 & 72.39 \\
\hline
ToyCar & 92.80 & 61.53 & 57.65 & 67.61 & 81.20 & 63.15 & 58.35 & 66.24 & 82.20 & 68.47 & 60.22 & 69.16 \\
\hline
ToyTrain & 88.38 & 28.49 & 52.15 & 45.74 & 81.77 & 31.72 & 52.89 & 47.88 & 84.62 & 32.98 & 52.42 & 49.00 \\
\hline
Valve & 51.43 & 51.00 & 50.50 & 50.97 & 71.56 & 60.47 & 60.99 & 63.96 & 72.34 & 65.65 & 62.15 & 66.45 \\
\hline
Overall & 74.89 & 48.87 & 56.27 & 58.15 & 75.97 & 54.17 & 59.25 & 61.85 & 78.98 & 56.80 & 60.75 & 64.20 \\
\hline
\end{tabular*}
\par
\end{table}

\begin{table}[!htbp]
\centering
\caption{Machine-wise $H$ for the separator configurations}
\label{tab:supp_separator_grid}
\footnotesize
\setlength{\arrayrulewidth}{0.3pt}
\setlength{\tabcolsep}{2.5pt}
\renewcommand{\arraystretch}{1.12}
\begin{tabular*}{\textwidth}{@{\extracolsep{\fill}}lrrrrrrrr@{}}
\hline
Configuration & Bearing & Fan & Gearbox & Slider & ToyCar & ToyTrain & Valve & Overall \\
\hline
\multicolumn{9}{@{}l}{\textit{GAP}} \\
\hline
0 blocks, 64 channels & 60.08 & 52.07 & 64.17 & 61.34 & 56.72 & 44.69 & 50.36 & 54.86 \\
\hline
0 blocks, 128 channels & 59.67 & 52.42 & 65.43 & 60.67 & 59.07 & 43.77 & 50.64 & 55.06 \\
\hline
1 block, 64 channels & 59.06 & 51.44 & 70.08 & 64.34 & 63.39 & 46.66 & 51.30 & 56.97 \\
\hline
1 block, 128 channels & 63.36 & 51.84 & 73.04 & 66.50 & 69.18 & 46.27 & 51.58 & 58.69 \\
\hline
2 blocks, 64 channels & 58.82 & 53.49 & 74.69 & 66.61 & 67.61 & 45.74 & 50.97 & 58.15 \\
\hline
2 blocks, 128 channels & 58.70 & 55.70 & 74.47 & 70.16 & 67.89 & 47.10 & 52.35 & 59.46 \\
\hline
3 blocks, 64 channels & 62.36 & 58.50 & 78.20 & 67.28 & 65.08 & 45.08 & 51.75 & 59.50 \\
\hline
3 blocks, 128 channels & 64.30 & 52.86 & 76.17 & 71.64 & 69.07 & 46.43 & 53.05 & 60.13 \\
\hline
4 blocks, 64 channels & 59.62 & 55.13 & 70.36 & 65.11 & 66.55 & 46.15 & 51.78 & 58.10 \\
\hline
4 blocks, 128 channels & 61.07 & 57.04 & 75.88 & 68.37 & 66.90 & 47.76 & 52.50 & 60.01 \\
\hline
\multicolumn{9}{@{}l}{\textit{Learned aggregation}} \\
\hline
0 blocks, 64 channels & 67.47 & 52.80 & 72.57 & 67.79 & 54.75 & 42.52 & 65.68 & 58.68 \\
\hline
0 blocks, 128 channels & 63.23 & 50.56 & 72.48 & 67.55 & 55.37 & 46.10 & 64.99 & 58.66 \\
\hline
1 block, 64 channels & 65.26 & 54.01 & 73.71 & 74.61 & 65.02 & 47.14 & 62.49 & 61.73 \\
\hline
1 block, 128 channels & 66.42 & 53.39 & 72.82 & 73.22 & 71.04 & 47.21 & 64.96 & 62.60 \\
\hline
2 blocks, 64 channels & 60.31 & 55.89 & 71.51 & 76.38 & 66.24 & 47.88 & 63.96 & 61.85 \\
\hline
2 blocks, 128 channels & 70.55 & 53.74 & 69.18 & 77.17 & 69.93 & 49.66 & 61.48 & 63.12 \\
\hline
3 blocks, 64 channels & 66.50 & 58.29 & 73.55 & 68.69 & 69.53 & 45.88 & 64.33 & 62.46 \\
\hline
3 blocks, 128 channels & 49.66 & 54.94 & 76.36 & 71.71 & 70.29 & 45.51 & 62.44 & 59.53 \\
\hline
4 blocks, 64 channels & 64.36 & 57.27 & 74.95 & 70.23 & 68.42 & 50.43 & 60.87 & 62.81 \\
\hline
4 blocks, 128 channels & 63.86 & 61.28 & 71.78 & 75.58 & 69.50 & 48.58 & 59.51 & 63.11 \\
\hline
\multicolumn{9}{@{}l}{\textit{Learned aggregation + correction}} \\
\hline
0 blocks, 64 channels & 65.18 & 53.13 & 72.59 & 70.13 & 60.29 & 44.42 & 57.59 & 58.99 \\
\hline
0 blocks, 128 channels & 63.08 & 52.95 & 73.88 & 74.52 & 62.46 & 48.53 & 59.21 & 60.77 \\
\hline
1 block, 64 channels & 63.89 & 55.41 & 78.58 & 73.63 & 68.05 & 49.67 & 61.59 & 63.02 \\
\hline
1 block, 128 channels & 67.31 & 58.02 & 76.77 & 71.78 & 72.11 & 46.50 & 63.47 & 63.53 \\
\hline
2 blocks, 64 channels & 68.60 & 57.31 & 75.33 & 72.39 & 69.16 & 49.00 & 66.45 & 64.20 \\
\hline
2 blocks, 128 channels & 69.22 & 57.38 & 73.09 & 74.49 & 67.98 & 49.78 & 62.31 & 63.74 \\
\hline
3 blocks, 64 channels & 69.90 & 58.54 & 74.57 & 70.00 & 71.79 & 48.02 & 65.32 & 64.13 \\
\hline
3 blocks, 128 channels & 62.08 & 56.22 & 74.84 & 72.58 & 71.02 & 47.36 & 64.44 & 62.64 \\
\hline
4 blocks, 64 channels & 54.97 & 57.56 & 74.27 & 70.78 & 71.41 & 47.73 & 63.27 & 61.45 \\
\hline
4 blocks, 128 channels & 66.62 & 58.90 & 74.22 & 74.27 & 71.29 & 49.38 & 62.65 & 64.12 \\
\hline
\end{tabular*}
\par
\end{table}

\begin{table}[!htbp]
\centering
\caption{Machine-wise $H$ for the separation training targets}
\label{tab:supp_objectives}
\footnotesize
\setlength{\arrayrulewidth}{0.3pt}
\setlength{\tabcolsep}{2.5pt}
\renewcommand{\arraystretch}{1.12}
\begin{tabular*}{\textwidth}{@{\extracolsep{\fill}}lrrrrrrrr@{}}
\hline
Configuration & Bearing & Fan & Gearbox & Slider & ToyCar & ToyTrain & Valve & Overall \\
\hline
\multicolumn{9}{@{}l}{\textit{2 blocks, 64 channels}} \\
\hline
Target, before correction & 60.31 & 55.89 & 71.51 & 76.38 & 66.24 & 47.88 & 63.96 & 61.85 \\
\hline
Non-target, before correction & 63.24 & 56.37 & 74.66 & 73.16 & 64.54 & 46.29 & 59.68 & 61.15 \\
\hline
Target, after correction & 68.60 & 57.31 & 75.33 & 72.39 & 69.16 & 49.00 & 66.45 & 64.20 \\
\hline
Non-target, after correction & 61.46 & 58.75 & 75.68 & 72.66 & 66.36 & 46.94 & 62.53 & 62.15 \\
\hline
\multicolumn{9}{@{}l}{\textit{3 blocks, 64 channels}} \\
\hline
Target, before correction & 66.50 & 58.29 & 73.55 & 68.69 & 69.53 & 45.88 & 64.33 & 62.46 \\
\hline
Non-target, before correction & 64.67 & 58.22 & 74.98 & 73.89 & 65.50 & 46.78 & 64.23 & 62.65 \\
\hline
Target, after correction & 69.90 & 58.54 & 74.57 & 70.00 & 71.79 & 48.02 & 65.32 & 64.13 \\
\hline
Non-target, after correction & 63.69 & 59.23 & 77.59 & 72.14 & 66.99 & 46.55 & 62.41 & 62.63 \\
\hline
\multicolumn{9}{@{}l}{\textit{4 blocks, 128 channels}} \\
\hline
Target, before correction & 63.86 & 61.28 & 71.78 & 75.58 & 69.50 & 48.58 & 59.51 & 63.11 \\
\hline
Non-target, before correction & 57.23 & 58.31 & 72.40 & 74.56 & 62.15 & 46.88 & 62.00 & 60.66 \\
\hline
Target, after correction & 66.62 & 58.90 & 74.22 & 74.27 & 71.29 & 49.38 & 62.65 & 64.12 \\
\hline
Non-target, after correction & 54.47 & 58.04 & 73.54 & 75.50 & 67.44 & 47.87 & 61.29 & 61.15 \\
\hline
\end{tabular*}
\par
\end{table}

\begin{table}[!htbp]
\centering
\caption{Machine-wise $H$ for the correction conditions}
\label{tab:supp_correction}
\footnotesize
\setlength{\arrayrulewidth}{0.3pt}
\setlength{\tabcolsep}{2.5pt}
\renewcommand{\arraystretch}{1.12}
\begin{tabular*}{\textwidth}{@{\extracolsep{\fill}}lrrrrrrrr@{}}
\hline
Correction & Bearing & Fan & Gearbox & Slider & ToyCar & ToyTrain & Valve & Overall \\
\hline
\multicolumn{9}{@{}l}{\textit{GAP}} \\
\hline
None & 58.82 & 53.49 & 74.69 & 66.61 & 67.61 & 45.74 & 50.97 & 58.15 \\
\hline
Mean centering & 60.37 & 56.68 & 67.97 & 63.29 & 77.44 & 45.16 & 50.89 & 58.61 \\
\hline
Interference covariance & 63.06 & 56.03 & 70.57 & 68.83 & 67.88 & 47.79 & 51.16 & 59.52 \\
\hline
Normal covariance + centering & 57.61 & 59.46 & 67.72 & 67.34 & 70.32 & 54.39 & 49.74 & 60.08 \\
\hline
Interference covariance + centering & 60.93 & 57.94 & 75.69 & 61.40 & 76.12 & 47.76 & 50.88 & 59.88 \\
\hline
\multicolumn{9}{@{}l}{\textit{Learned aggregation}} \\
\hline
None & 60.31 & 55.89 & 71.51 & 76.38 & 66.24 & 47.88 & 63.96 & 61.85 \\
\hline
Mean centering & 65.69 & 57.25 & 70.04 & 74.45 & 65.94 & 48.32 & 64.76 & 62.67 \\
\hline
Interference covariance & 67.13 & 55.99 & 75.53 & 72.34 & 65.17 & 49.50 & 59.36 & 62.38 \\
\hline
Normal covariance + centering & 58.40 & 55.51 & 69.55 & 71.46 & 69.58 & 51.21 & 53.53 & 60.30 \\
\hline
Interference covariance + centering & 68.60 & 57.31 & 75.33 & 72.39 & 69.16 & 49.00 & 66.45 & 64.20 \\
\hline
\end{tabular*}
\par
\end{table}

\begin{table}[!htbp]
\centering
\caption{Machine-wise $H$ for the scorer conditions}
\label{tab:supp_scorers}
\footnotesize
\setlength{\arrayrulewidth}{0.3pt}
\setlength{\tabcolsep}{1.4pt}
\renewcommand{\arraystretch}{1.12}
\begin{minipage}[t]{0.49\textwidth}
\centering
(a) GAP\\[3pt]
\begin{tabular*}{\linewidth}{@{\extracolsep{\fill}}lrrrrrrrr@{}}
\hline
Scorer & Bea. & Fan & Gear. & Sli. & Car & Train & Val. & All \\
\hline
Cos. 1-NN & 58.82 & 53.49 & 74.69 & 66.61 & 67.61 & 45.74 & 50.97 & 58.15 \\
\hline
Cos. 5-NN & 57.51 & 53.61 & 74.38 & 64.84 & 61.67 & 43.78 & 50.57 & 56.57 \\
\hline
MD-Emp & 53.22 & 54.32 & 69.96 & 63.81 & 50.43 & 43.82 & 51.27 & 54.13 \\
\hline
MD-LW & 52.40 & 55.05 & 71.78 & 62.99 & 50.75 & 43.80 & 51.68 & 54.29 \\
\hline
Density & 54.93 & 55.42 & 65.53 & 69.16 & 68.73 & 59.21 & 51.57 & 59.93 \\
\hline
\end{tabular*}
\end{minipage}\hfill
\begin{minipage}[t]{0.49\textwidth}
\centering
(b) GAP + correction\\[3pt]
\begin{tabular*}{\linewidth}{@{\extracolsep{\fill}}lrrrrrrrr@{}}
\hline
Scorer & Bea. & Fan & Gear. & Sli. & Car & Train & Val. & All \\
\hline
Cos. 1-NN & 60.93 & 57.94 & 75.69 & 61.40 & 76.12 & 47.76 & 50.88 & 59.88 \\
\hline
Cos. 5-NN & 61.21 & 57.69 & 75.72 & 57.64 & 72.71 & 45.01 & 50.04 & 58.25 \\
\hline
MD-Emp & 53.22 & 54.32 & 69.96 & 63.81 & 50.43 & 43.82 & 51.27 & 54.13 \\
\hline
MD-LW & 53.21 & 54.35 & 70.02 & 63.73 & 50.49 & 43.98 & 51.29 & 54.18 \\
\hline
Density & 49.30 & 56.54 & 67.20 & 66.82 & 73.36 & 51.78 & 53.46 & 58.59 \\
\hline
\end{tabular*}
\end{minipage}

\vspace{8pt}
\begin{minipage}[t]{0.49\textwidth}
\centering
(c) Learned aggregation\\[3pt]
\begin{tabular*}{\linewidth}{@{\extracolsep{\fill}}lrrrrrrrr@{}}
\hline
Scorer & Bea. & Fan & Gear. & Sli. & Car & Train & Val. & All \\
\hline
Cos. 1-NN & 60.31 & 55.89 & 71.51 & 76.38 & 66.24 & 47.88 & 63.96 & 61.85 \\
\hline
Cos. 5-NN & 58.67 & 57.16 & 70.38 & 74.52 & 61.62 & 47.15 & 62.16 & 60.51 \\
\hline
MD-Emp & 52.08 & 55.73 & 71.84 & 61.19 & 49.01 & 41.99 & 40.29 & 51.30 \\
\hline
MD-LW & 48.60 & 55.93 & 71.82 & 57.47 & 47.82 & 41.96 & 37.88 & 49.67 \\
\hline
Density & 56.89 & 43.69 & 66.11 & 69.45 & 70.30 & 51.31 & 72.24 & 59.55 \\
\hline
\end{tabular*}
\end{minipage}\hfill
\begin{minipage}[t]{0.49\textwidth}
\centering
(d) Learned aggregation + correction\\[3pt]
\begin{tabular*}{\linewidth}{@{\extracolsep{\fill}}lrrrrrrrr@{}}
\hline
Scorer & Bea. & Fan & Gear. & Sli. & Car & Train & Val. & All \\
\hline
Cos. 1-NN & 68.60 & 57.31 & 75.33 & 72.39 & 69.16 & 49.00 & 66.45 & 64.20 \\
\hline
Cos. 5-NN & 67.86 & 58.59 & 74.30 & 71.63 & 64.29 & 47.18 & 64.71 & 62.82 \\
\hline
MD-Emp & 52.08 & 55.73 & 71.84 & 61.19 & 49.01 & 41.99 & 40.29 & 51.30 \\
\hline
MD-LW & 51.87 & 55.74 & 71.81 & 60.96 & 48.96 & 41.90 & 39.95 & 51.14 \\
\hline
Density & 55.72 & 49.41 & 71.37 & 74.08 & 72.45 & 52.53 & 73.48 & 62.42 \\
\hline
\end{tabular*}
\end{minipage}
\par\smallskip
\parbox{\textwidth}{\footnotesize Bea.: bearing; Gear.: gearbox; Sli.: slider; Car: ToyCar; Train: ToyTrain; Val.: valve; All: overall. Density denotes local density-based cosine normalization.}
\par
\end{table}

\begin{table}[!htbp]
\centering
\caption{Machine-wise metrics for autoencoder, CE, and ArcFace}
\label{tab:supp_comparisons_a}
\footnotesize
\setlength{\arrayrulewidth}{0.3pt}
\setlength{\tabcolsep}{2.4pt}
\renewcommand{\arraystretch}{1.12}
\begin{tabular*}{\textwidth}{@{\extracolsep{\fill}}lrrrrrrrrrrrr@{}}
\hline
Machine & \multicolumn{4}{c}{Autoencoder} & \multicolumn{4}{c}{CE} & \multicolumn{4}{c}{ArcFace} \\
\cline{2-5} \cline{6-9} \cline{10-13}
 & Src. & Tgt. & pAUC & $H$ & Src. & Tgt. & pAUC & $H$ & Src. & Tgt. & pAUC & $H$ \\
\hline
Bearing & 61.64 & 67.38 & 52.99 & 60.07 & 60.72 & 49.83 & 56.09 & 55.18 & 64.03 & 42.16 & 55.68 & 52.36 \\
\hline
Fan & 80.00 & 49.57 & 57.90 & 60.07 & 81.03 & 54.05 & 57.47 & 62.19 & 80.42 & 45.98 & 57.80 & 58.27 \\
\hline
Gearbox & 74.21 & 67.13 & 60.69 & 66.89 & 68.63 & 55.29 & 57.43 & 59.92 & 59.88 & 50.71 & 53.94 & 54.59 \\
\hline
Slider & 79.04 & 48.49 & 56.74 & 58.94 & 95.94 & 80.84 & 75.18 & 83.12 & 89.66 & 79.52 & 75.76 & 81.24 \\
\hline
ToyCar & 90.31 & 48.64 & 53.85 & 59.76 & 62.65 & 55.18 & 54.19 & 57.10 & 72.67 & 49.23 & 53.94 & 57.02 \\
\hline
ToyTrain & 77.41 & 33.00 & 50.96 & 47.73 & 72.53 & 38.35 & 52.63 & 50.97 & 59.03 & 34.35 & 50.29 & 45.50 \\
\hline
Valve & 53.66 & 49.91 & 50.56 & 51.32 & 66.62 & 72.38 & 59.81 & 65.87 & 73.44 & 76.43 & 63.94 & 70.86 \\
\hline
Overall & 71.81 & 49.49 & 54.59 & 57.20 & 71.02 & 55.10 & 58.28 & 60.74 & 69.84 & 49.90 & 57.82 & 58.08 \\
\hline
\end{tabular*}
\par
\end{table}

\begin{table}[!htbp]
\centering
\caption{Machine-wise metrics for SimCLR, SimSiam, and direct paired-mixture learning}
\label{tab:supp_comparisons_b}
\footnotesize
\setlength{\arrayrulewidth}{0.3pt}
\setlength{\tabcolsep}{2.4pt}
\renewcommand{\arraystretch}{1.12}
\begin{tabular*}{\textwidth}{@{\extracolsep{\fill}}lrrrrrrrrrrrr@{}}
\hline
Machine & \multicolumn{4}{c}{SimCLR} & \multicolumn{4}{c}{SimSiam} & \multicolumn{4}{c}{Direct paired-mixture} \\
\cline{2-5} \cline{6-9} \cline{10-13}
 & Src. & Tgt. & pAUC & $H$ & Src. & Tgt. & pAUC & $H$ & Src. & Tgt. & pAUC & $H$ \\
\hline
Bearing & 61.64 & 46.02 & 51.67 & 52.35 & 47.94 & 55.03 & 49.60 & 50.68 & 53.28 & 66.76 & 55.18 & 57.84 \\
\hline
Fan & 75.10 & 46.79 & 57.26 & 57.53 & 69.48 & 54.55 & 57.24 & 59.77 & 77.75 & 45.57 & 59.89 & 58.25 \\
\hline
Gearbox & 62.93 & 56.28 & 53.15 & 57.17 & 63.09 & 59.79 & 56.89 & 59.82 & 70.43 & 75.17 & 59.60 & 67.75 \\
\hline
Slider & 78.23 & 59.91 & 53.02 & 62.07 & 64.53 & 55.01 & 52.99 & 57.09 & 98.76 & 64.25 & 65.78 & 73.36 \\
\hline
ToyCar & 69.46 & 52.00 & 54.12 & 57.58 & 87.29 & 41.90 & 52.69 & 55.25 & 69.21 & 61.91 & 57.94 & 62.68 \\
\hline
ToyTrain & 67.97 & 35.52 & 50.92 & 48.00 & 76.14 & 33.88 & 52.17 & 48.53 & 72.67 & 48.89 & 51.12 & 55.79 \\
\hline
Valve & 55.86 & 55.99 & 53.07 & 54.94 & 53.91 & 53.45 & 51.55 & 52.95 & 81.27 & 72.99 & 71.12 & 74.87 \\
\hline
Overall & 66.53 & 49.02 & 53.25 & 55.34 & 63.79 & 48.78 & 53.18 & 54.56 & 72.56 & 60.29 & 59.48 & 63.59 \\
\hline
\end{tabular*}
\par
\end{table}

\begin{table}[!htbp]
\centering
\caption{Machine-wise metrics for separation with GAP (three blocks, 128 channels), separation with aggregation (two blocks, 128 channels), and BEATs iter3 with a fixed frequency-concatenated readout}
\label{tab:supp_comparisons_c}
\footnotesize
\setlength{\arrayrulewidth}{0.3pt}
\setlength{\tabcolsep}{2.4pt}
\renewcommand{\arraystretch}{1.12}
\begin{tabular*}{\textwidth}{@{\extracolsep{\fill}}lrrrrrrrrrrrr@{}}
\hline
Machine & \multicolumn{4}{c}{Separation + GAP} & \multicolumn{4}{c}{Separation + aggregation} & \multicolumn{4}{c}{BEATs} \\
\cline{2-5} \cline{6-9} \cline{10-13}
 & Src. & Tgt. & pAUC & $H$ & Src. & Tgt. & pAUC & $H$ & Src. & Tgt. & pAUC & $H$ \\
\hline
Bearing & 69.98 & 63.56 & 60.12 & 64.30 & 69.70 & 85.10 & 60.88 & 70.55 & 66.83 & 70.84 & 59.24 & 65.27 \\
\hline
Fan & 74.79 & 40.75 & 53.08 & 52.86 & 81.25 & 40.39 & 53.33 & 53.74 & 77.81 & 54.99 & 62.73 & 63.86 \\
\hline
Gearbox & 81.57 & 78.38 & 69.60 & 76.17 & 74.94 & 71.95 & 62.03 & 69.18 & 78.88 & 69.07 & 62.16 & 69.38 \\
\hline
Slider & 90.08 & 68.84 & 61.55 & 71.64 & 97.28 & 72.97 & 67.15 & 77.17 & 89.29 & 76.78 & 65.03 & 75.75 \\
\hline
ToyCar & 91.54 & 65.68 & 57.85 & 69.07 & 85.37 & 67.57 & 61.02 & 69.93 & 91.63 & 74.06 & 55.89 & 70.91 \\
\hline
ToyTrain & 89.16 & 29.46 & 51.40 & 46.43 & 81.91 & 33.82 & 53.67 & 49.66 & 86.47 & 34.16 & 51.94 & 49.92 \\
\hline
Valve & 53.11 & 54.78 & 51.38 & 53.05 & 64.70 & 61.77 & 58.30 & 61.48 & 55.29 & 52.16 & 50.14 & 52.45 \\
\hline
Overall & 76.10 & 51.86 & 57.24 & 60.13 & 78.08 & 56.15 & 59.14 & 63.12 & 75.87 & 57.49 & 57.67 & 62.61 \\
\hline
\end{tabular*}
\par
\end{table}

\begin{table}[!htbp]
\centering
\caption{Machine-wise metrics for EAT, CED, and Dis-multi-PL}
\label{tab:supp_comparisons_d}
\footnotesize
\setlength{\arrayrulewidth}{0.3pt}
\setlength{\tabcolsep}{2.4pt}
\renewcommand{\arraystretch}{1.12}
\begin{tabular*}{\textwidth}{@{\extracolsep{\fill}}lrrrrrrrrrrrr@{}}
\hline
Machine & \multicolumn{4}{c}{EAT} & \multicolumn{4}{c}{CED} & \multicolumn{4}{c}{Dis-multi-PL} \\
\cline{2-5} \cline{6-9} \cline{10-13}
 & Src. & Tgt. & pAUC & $H$ & Src. & Tgt. & pAUC & $H$ & Src. & Tgt. & pAUC & $H$ \\
\hline
Bearing & 62.65 & 60.73 & 51.99 & 58.07 & 63.88 & 49.49 & 51.15 & 54.14 & 66.73 & 65.70 & 58.49 & 63.42 \\
\hline
Fan & 81.20 & 52.32 & 63.65 & 63.64 & 74.80 & 46.57 & 55.55 & 56.77 & 70.22 & 54.52 & 59.07 & 60.59 \\
\hline
Gearbox & 81.95 & 67.79 & 60.51 & 69.00 & 72.83 & 58.72 & 52.89 & 60.40 & 84.80 & 72.93 & 64.01 & 72.94 \\
\hline
Slider & 85.04 & 66.31 & 54.16 & 66.22 & 84.92 & 61.19 & 54.23 & 64.43 & 96.68 & 74.21 & 72.79 & 79.88 \\
\hline
ToyCar & 92.71 & 67.66 & 58.60 & 70.37 & 76.18 & 47.29 & 51.07 & 55.70 & 84.73 & 75.12 & 59.15 & 71.39 \\
\hline
ToyTrain & 83.89 & 31.80 & 52.49 & 48.06 & 83.71 & 50.36 & 53.80 & 59.53 & 81.91 & 37.99 & 52.67 & 52.16 \\
\hline
Valve & 55.49 & 55.86 & 51.17 & 54.09 & 59.29 & 54.98 & 51.25 & 54.98 & 73.16 & 63.49 & 54.14 & 62.64 \\
\hline
Overall & 75.30 & 54.04 & 55.74 & 60.33 & 72.56 & 52.14 & 52.80 & 57.80 & 78.62 & 60.27 & 59.45 & 65.03 \\
\hline
\end{tabular*}
\par
\end{table}

\begin{table}[!htbp]
\centering
\caption{Configuration-level $H$ for autoencoder and direct paired-mixture learning}
\label{tab:supp_ae_direct}
\footnotesize
\setlength{\arrayrulewidth}{0.3pt}
\begin{minipage}[t]{0.46\textwidth}
\centering
(a) Autoencoder with reconstruction MSE\\[3pt]
\begin{tabular*}{\linewidth}{@{\extracolsep{\fill}}crrr@{}}
\hline
Hidden & \multicolumn{3}{c}{Latent dimension} \\
\cline{2-4}
 & 4 & 8 & 16 \\
\hline
64 & 53.39 & 52.82 & 52.30 \\
\hline
128 & 54.06 & 54.51 & 54.77 \\
\hline
256 & 55.28 & 56.17 & 57.20 \\
\hline
\end{tabular*}
\end{minipage}\hfill
\begin{minipage}[t]{0.50\textwidth}
\centering
(b) Direct paired-mixture learning with cosine 1-NN\\[3pt]
\begin{tabular*}{\linewidth}{@{\extracolsep{\fill}}lrr@{}}
\hline
Backbone & Before correction & After correction \\
\hline
ResNet-18 & 63.59 & 63.10 \\
\hline
ResNet-34 & 58.83 & 59.35 \\
\hline
ResNet-50 & 57.86 & 59.28 \\
\hline
ResNet-101 & 58.27 & 58.85 \\
\hline
ResNet-152 & 56.27 & 54.86 \\
\hline
\end{tabular*}
\end{minipage}
\par
\end{table}

\begin{table}[!htbp]
\centering
\caption{Configuration-level $H$ for classification and self-supervised ResNets}
\label{tab:supp_resnet}
\footnotesize
\setlength{\arrayrulewidth}{0.3pt}
\setlength{\tabcolsep}{3.0pt}
\renewcommand{\arraystretch}{1.12}
\begin{tabular*}{\textwidth}{@{\extracolsep{\fill}}lrrrrrr@{}}
\hline
Backbone & CE & ArcFace & \multicolumn{2}{c}{SimCLR} & \multicolumn{2}{c}{SimSiam} \\
\cline{4-5}\cline{6-7}
 &  &  & Backbone & Projection & Backbone & Projection \\
\hline
ResNet-18 & 58.88 & 57.65 & 54.65 & 54.58 & 54.02 & 53.63 \\
\hline
ResNet-34 & 60.34 & 57.23 & 54.77 & 55.34 & 52.74 & 52.98 \\
\hline
ResNet-50 & 60.74 & 57.96 & 54.21 & 52.97 & 54.53 & 52.52 \\
\hline
ResNet-101 & 57.86 & 57.70 & 53.91 & 53.32 & 52.36 & 49.71 \\
\hline
ResNet-152 & 57.76 & 58.08 & 53.99 & 54.24 & 54.56 & 51.24 \\
\hline
\end{tabular*}
\par
\end{table}

\begin{table}[!htbp]
\centering
\caption{Frozen pre-trained feature configurations and overall $H$}
\label{tab:supp_pretrained}
\footnotesize
\setlength{\arrayrulewidth}{0.3pt}
\setlength{\tabcolsep}{2.5pt}
\renewcommand{\arraystretch}{1.12}
\begin{tabular*}{\textwidth}{@{\extracolsep{\fill}}llrr@{}}
\hline
Model / checkpoint & Feature construction & Dimension & $H$ \\
\hline
BEATs iter3 & Patch mean & 768 & 60.81 \\
\hline
BEATs iter3 & Time mean + frequency concatenation & 6144 & 62.61 \\
\hline
BEATs iter3+ (AS2M) & Patch mean & 768 & 60.26 \\
\hline
BEATs iter3+ (AS2M) & Time mean + frequency concatenation & 6144 & 60.47 \\
\hline
EAT-base (epoch 30) & CLS & 768 & 58.66 \\
\hline
EAT-large (epoch 20) & CLS & 1024 & 59.24 \\
\hline
EAT-large (epoch 20) & CLS + time mean + frequency concatenation & 9216 & 60.33 \\
\hline
CED-tiny & Encoder patch mean & 192 & 56.88 \\
\hline
CED-mini & Encoder patch mean & 256 & 57.38 \\
\hline
CED-small & Encoder patch mean & 384 & 57.80 \\
\hline
CED-base & Encoder patch mean & 768 & 57.68 \\
\hline
\end{tabular*}
\par
\end{table}

\FloatBarrier

\section{Implementation and Reproducibility Details}
\label{supp:implementation}

\subsection{CMGAN-Based System}

Table~\ref{tab:implementation_details} lists the settings of the two-block, 64-channel system. Each feature stage has independent adapter, value-projection, query, and linear-layer parameters. Region boundaries are integer floors of proportional time and frequency positions. Region vectors are concatenated in time-major, frequency-minor order, and stage vectors follow encoder-to-block order before fusion.

\begin{table}[!htbp]
\centering
\caption{Implementation details of the representative system}
\label{tab:implementation_details}
\footnotesize
\setlength{\arrayrulewidth}{0.3pt}
\setlength{\tabcolsep}{3pt}
\renewcommand{\arraystretch}{1.12}
\begin{tabular}{@{}>{\raggedright\arraybackslash}p{0.39\columnwidth}>{\raggedright\arraybackslash}p{0.56\columnwidth}@{}}
\hline
Component & Setting \\
\hline
Feature stages & Encoder and two Conformer outputs \\
\hline
Channels per stage & 64 \\
\hline
Adapter depthwise convolution & $3\times3$ kernel, stride 1, padding 1, 64 groups \\
\hline
Adapter pointwise convolution & $1\times1$, $64\rightarrow64$ \\
\hline
Value projection & $1\times1$, $64\rightarrow64$ \\
\hline
Region queries per stage & Eight 64-dimensional vectors \\
\hline
Region-to-stage projection & Linear, $512\rightarrow128$ \\
\hline
Stage fusion & Linear, $384\rightarrow128$ \\
\hline
Training-only projection & $128\rightarrow256\rightarrow128$, with GELU between the layers \\
\hline
Convolution and linear biases & Enabled \\
\hline
Separator / aggregation seeds (original run) & 0 / 1 \\
\hline
Calibration seed & 0 \\
\hline
Calibration samples & 512 recordings, 2,560 segment differences \\
\hline
Reference bank & 29,995 segment embeddings for bearing and 30,000 for each other type \\
\hline
Covariance computation & CPU, double precision \\
\hline
Covariance eigenvalue floor & $10^{-8}$ times the largest eigenvalue \\
\hline
Empirical MD pseudoinverse threshold & $10^{-12}$ relative to the largest singular value \\
\hline
Local density scale floor & $10^{-12}$ \\
\hline
\end{tabular}
\end{table}

Each calibration recording supplies one pair of mixture views and five aligned segment differences. Ledoit--Wolf estimation centers the differences by their own mean. Covariance eigenvalues are floored as specified in Table~\ref{tab:implementation_details} before computing the symmetric inverse square root. The centering mean uses the full original normal reference bank, and correction is applied to inputs and references. Calibration recordings belong to the normal data used for representation learning.

Empirical Mahalanobis scoring uses sample covariance with divisor $N_{\mathrm{bank}}-1$. For a full-rank covariance matrix with a condition number below $10^{12}$, the scorer uses an ordinary inverse. For other matrices it uses the Moore--Penrose pseudoinverse with a relative singular-value threshold of $10^{-12}$. The Ledoit--Wolf scorer uses its fitted shrinkage precision. Both return the square root of the nonnegative quadratic form. For CMGAN, local density normalization uses the five nearest normal segments from other original recordings and a scale floor of $10^{-12}$. Recording scores average the five segment scores.

\subsection{Comparison Systems}

The autoencoder and ResNet baselines use 128-bin log-mel spectrograms with a 1,024-point FFT and a 512-sample hop. The autoencoder concatenates five frames into a 640-dimensional input and uses the hidden and latent dimensions in Table~\ref{tab:supp_ae_direct}. It is trained for 200 epochs using Adam with a learning rate of $10^{-3}$ and a batch size of 512 frame groups.

The classification and self-supervised ResNets in Table~\ref{tab:supp_resnet} are randomly initialized and process original 10-s recordings. Training uses AdamW for 200 epochs with a batch size of 64 and an initial learning rate of $5\times10^{-4}$, halved every 50 epochs. SimCLR and SimSiam use gain changes and time--frequency masking. Classification features are extracted before the final layer. SimCLR and SimSiam use separate backbone and projection readouts.

The direct paired-mixture ResNets in Table~\ref{tab:supp_ae_direct} are trained separately for each target type using the paired mixtures sharing a target recording and the contrastive and target-reference consistency objectives. The backbone features from each 2-s log-mel input are mapped to 128-dimensional embeddings by a linear layer. The backbone, embedding layer, and projection are jointly trained for 200 epochs using AdamW with a learning rate of $10^{-3}$ and a batch size of 16 target recordings. Each model is evaluated before and after correction.

Features from the frozen BEATs, EAT, and CED models are extracted using the readouts in Table~\ref{tab:supp_pretrained} and each model's released preprocessing. The Dis-multi-PL student is trained for 16 epochs using AdamW with a batch size of 64, a learning rate of $10^{-3}$, and weight decay of 0.01.

\section{Computational Complexity}
\label{supp:complexity}

Table~\ref{tab:supp_complexity} reports the parameter counts and multiply--accumulate operations (MACs) of the listed CMGAN and comparison configurations. The BEATs, EAT, and CED entries use fixed readouts. Parameter counts include frozen weights required for ASD and refer to one machine-specific model where applicable. MACs are counted for one native network input, as specified in the table. MACs cover convolution, linear, and attention products, including regional aggregation. Signal preprocessing, activation and normalization operations, and normal-reference search are excluded.

\begin{table}[!htbp]
\centering
\caption{Parameters and MACs of the representative ASD configurations}
\label{tab:supp_complexity}
\small
\setlength{\arrayrulewidth}{0.3pt}
\setlength{\tabcolsep}{3pt}
\renewcommand{\arraystretch}{1.18}
\begin{tabular*}{\textwidth}{@{\extracolsep{\fill}}lllrr@{}}
\hline
Method & Configuration & Input & Parameters (M) & MACs (G) \\
\hline
\multicolumn{5}{@{}l}{\textit{Without externally pre-trained models}} \\
\hline
Autoencoder & Hidden 256 / latent 16 & 5 frames & 0.736 & $7.29\times10^{-4}$ \\
\hline
CE & ResNet-50 & 10 s & 23.502 & 3.264 \\
\hline
ArcFace & ResNet-152 & 10 s & 58.138 & 9.325 \\
\hline
SimCLR & ResNet-34, projected & 10 s & 21.871 & 2.919 \\
\hline
SimSiam & ResNet-152 & 10 s & 58.138 & 9.325 \\
\hline
Direct paired-mixture & ResNet-18 & 2 s & 11.236 & 0.283 \\
\hline
Separation + GAP & 3 blocks / 128 ch & 2 s & 3.861 & 79.868 \\
\hline
Separation + aggregation & 2 blocks / 128 ch & 2 s & 3.245 & 65.381 \\
\hline
\textbf{CMGAN + regional + correction} & 2 blocks / 64 ch & 2 s & 1.050 & 17.414 \\
\hline
\multicolumn{5}{@{}l}{\textit{With externally pre-trained models}} \\
\hline
BEATs & iter3, freq. concat. & 10 s & 90.312 & 49.304 \\
\hline
EAT & Large, CLS + freq. concat. & 10 s & 308.867 & 165.172 \\
\hline
CED & Small, patch mean & 10 s & 21.419 & 5.857 \\
\hline
Dis-multi-PL & Student embedding & 10 s & 5.000 & 1.156 \\
\hline
\end{tabular*}
\par\smallskip
\parbox{\textwidth}{\footnotesize M denotes $10^6$ parameters and G denotes $10^9$ MACs. The counts include the autoencoder reconstruction decoder and the SimCLR projection used for scoring. Separation decoders and heads unused for ASD are excluded. Dis-multi-PL counts the student extractor, with the BEATs pseudo-label teacher excluded. The proposed correction adds $128^2=16{,}384$ MACs per embedding. Its fitted mean and matrix are excluded from neural-network parameter counts. The table gives MACs for each system's native input unit. A 10-s recording requires five 2-s passes for the separation and direct paired-mixture systems.}
\end{table}

\FloatBarrier
\section{Scoring of fixed CMGAN features}
\label{supp:scoring}
Table~\ref{tab:local_scorer_comparison} compares five scoring rules on the same recordings from sections 00--02 and frozen two-block, 64-channel separator. The four columns use global average pooling (GAP) or the learned regional readout before and after correction. Table~\ref{tab:supp_scorers} gives the machine-level results for each combination. Cosine 5-NN averages the distances to five reference segments. Local density normalization divides each input-to-reference cosine distance by the sum of the distances from that reference to its five nearest normal segments from other recordings, then takes the minimum normalized distance.

\begin{table}[!htbp]
\centering
\caption{Scorer comparison for the two-block, 64-channel CMGAN system. Values are overall $H$ (\%).}
\label{tab:local_scorer_comparison}
\footnotesize
\setlength{\arrayrulewidth}{0.3pt}
\setlength{\tabcolsep}{2.2pt}
\renewcommand{\arraystretch}{1.12}
\begin{tabular*}{\linewidth}{@{\extracolsep{\fill}}lrrrr@{}}
\hline
Scorer & GAP & \shortstack{GAP\\+ corr.} & Learned & \shortstack{Learned\\+ corr.} \\
\hline
Cosine 1-NN & 58.15 & 59.88 & 61.85 & 64.20 \\
\hline
Cosine 5-NN & 56.57 & 58.25 & 60.51 & 62.82 \\
\hline
MD-Emp & 54.13 & 54.13 & 51.30 & 51.30 \\
\hline
MD-LW & 54.29 & 54.18 & 49.67 & 51.14 \\
\hline
Density & 59.93 & 58.59 & 59.55 & 62.42 \\
\hline
\end{tabular*}
\smallskip
\parbox{\linewidth}{\footnotesize Corr.\ denotes centering and interference-covariance correction. MD-Emp and MD-LW use empirical and Ledoit--Wolf covariance, respectively. Density denotes local density-based cosine normalization.}
\end{table}

The recording-level scatterplots in the main article compare MD-LW and cosine 1-NN for the same corrected CMGAN embeddings.

\FloatBarrier
\section{Frozen encoders and training variability}
\label{supp:encoder}
Table~\ref{tab:proxy_regional_detection} gives all task-trained frozen encoders used to evaluate the regional readout. Each ResNet contributes its final convolutional map before pooling, partitioned into two temporal by four frequency regions. The autoencoder (AE) contributes hidden features from its fourth encoder block before the bottleneck, divided into eight temporal regions. Each model produces one embedding per 10-s recording. The mean readout pools the same features as the regional readout. The reconstruction-MSE, latent and residual AE baselines in the other comparisons use distinct outputs. The encoder weights and normalization statistics remain fixed. For each encoder and machine type, the aggregation head and training projection are trained for 20 epochs on normal recordings, with the checkpoint selected by the lowest epoch-average training loss. The corrected and uncorrected regional readouts share this checkpoint.

\begin{table}[!htbp]
    \centering
    \caption{Overall $H$ (\%) on sections 00--02 for regional aggregation applied to encoders trained on the normal task data. Mean pools the same feature map or hidden sequence used by Regional. Regional + correction uses the same aggregation checkpoint as Regional. Results use cosine 1-NN scoring. Bold numbers indicate the highest value in each row.}
    \label{tab:proxy_regional_detection}

    \small
    \linespread{1.0}\selectfont
    \renewcommand{\arraystretch}{1.10}
    \setlength{\tabcolsep}{4pt}
    \begin{tabular*}{\linewidth}{@{\extracolsep{\fill}}llrrr@{}}
        \toprule
        \textbf{Frozen encoder} & \textbf{Training objective} & \textbf{Mean} & \textbf{Regional} & \shortstack{\textbf{Regional}\\\textbf{+ correction}} \\
        \midrule
        ResNet-18 & CE & 58.88 & \textbf{60.04} & 59.77 \\
        ResNet-18 & ArcFace & 57.65 & \textbf{58.86} & 57.01 \\
        ResNet-18 & SimCLR & 54.65 & \textbf{57.96} & 57.77 \\
        ResNet-18 & SimSiam & 54.02 & 52.78 & \textbf{55.23} \\
        \addlinespace[3pt]
        ResNet-34 & CE & \textbf{60.34} & 58.57 & 59.47 \\
        ResNet-34 & ArcFace & 57.23 & 56.07 & \textbf{57.79} \\
        ResNet-34 & SimCLR & 54.77 & \textbf{58.88} & 58.29 \\
        ResNet-34 & SimSiam & 52.74 & 51.75 & \textbf{53.91} \\
        \addlinespace[3pt]
        ResNet-50 & CE & \textbf{60.74} & 58.09 & 57.87 \\
        ResNet-50 & ArcFace & \textbf{57.96} & 54.08 & 52.42 \\
        ResNet-50 & SimCLR & 54.21 & \textbf{58.76} & 57.95 \\
        ResNet-50 & SimSiam & \textbf{54.53} & 52.90 & 50.36 \\
        \addlinespace[3pt]
        ResNet-101 & CE & \textbf{57.86} & 55.80 & 56.09 \\
        ResNet-101 & ArcFace & \textbf{57.70} & 54.59 & 50.95 \\
        ResNet-101 & SimCLR & 53.91 & \textbf{58.99} & 56.57 \\
        ResNet-101 & SimSiam & 52.36 & 51.64 & \textbf{54.05} \\
        \addlinespace[3pt]
        ResNet-152 & CE & \textbf{57.76} & 56.41 & 55.55 \\
        ResNet-152 & ArcFace & \textbf{58.08} & 55.82 & 53.82 \\
        ResNet-152 & SimCLR & 53.99 & \textbf{58.64} & 57.61 \\
        ResNet-152 & SimSiam & 54.56 & 55.92 & \textbf{56.88} \\
        \addlinespace[3pt]
        AE H128/Z8 & Reconstruction & 55.47 & 56.35 & \textbf{56.93} \\
        \bottomrule
    \end{tabular*}
\end{table}

Regional aggregation improved the uncorrected SimCLR features at all five backbone sizes. The effects varied with the other training objectives. For example, mean pooling on ResNet-50 retained higher $H$ for both CE and ArcFace than the regional readout in either correction condition.

\FloatBarrier
The repeated-run comparison in Table~\ref{tab:appendix_encoder_repeats} uses the original development-test recordings. In run order, CMGAN uses separator seeds 0, 1 and 2 with aggregation seeds 1, 2 and 3. BEATs uses the same frozen iter3 encoder with aggregation seeds 1, 2 and 3. Dis-multi-PL uses clustering seeds 42, 43 and 44 and student-training seeds 0, 1 and 2. Direct paired-mixture ResNet-18 uses seeds 1, 2 and 3 and is scored before correction. The SNR tables in Section~\ref{supp:noise} report one result per condition.

\begin{table}[!htbp]
    \centering
    \caption{Overall $H$ (\%) over three training runs. Mean $\pm$ SD uses the arithmetic mean and sample standard deviation.}
    \label{tab:appendix_encoder_repeats}
    \small
    \linespread{1.0}\selectfont
    \renewcommand{\arraystretch}{1.15}
    \setlength{\tabcolsep}{4pt}
    \begin{tabular*}{\linewidth}{@{\extracolsep{\fill}}lrrrr@{}}
        \toprule
        \textbf{System} & \textbf{Run 1} & \textbf{Run 2} & \textbf{Run 3} & \textbf{Mean $\pm$ SD} \\
        \midrule
        CMGAN + regional + correction & 64.20 & 64.86 & 64.62 & $64.56\pm0.33$ \\
        BEATs + regional + correction & 65.35 & 65.12 & 64.17 & $64.88\pm0.62$ \\
        Direct paired-mixture (ResNet-18) & 63.59 & 62.66 & 60.74 & $62.33\pm1.45$ \\
        Dis-multi-PL & 65.03 & 65.66 & 63.93 & $64.87\pm0.87$ \\
        \bottomrule
    \end{tabular*}
\end{table}

Table~\ref{tab:beats_domain_detection} reports the first BEATs regional run by machine and domain. Table~\ref{tab:supp_comparisons_c} gives the corresponding fixed-readout BEATs results, and Table~\ref{tab:supp_proposed_metrics} contains the CMGAN results.

\begin{table}[!htbp]
    \centering
    \caption{Machine- and domain-wise performance (\%) on sections 00--02 with frozen BEATs iter3, regional aggregation, and embedding correction.}
    \label{tab:beats_domain_detection}
    \small
    \linespread{1.0}\selectfont
    \renewcommand{\arraystretch}{1.15}
    \setlength{\tabcolsep}{5pt}
    \begin{tabular*}{\linewidth}{@{\extracolsep{\fill}}lrrrr@{}}
        \toprule
        \textbf{Machine type} & \textbf{Source AUC} & \textbf{Target AUC} & \textbf{pAUC} & $H$ \\
        \midrule
        Bearing & 61.29 & 73.58 & 54.44 & 62.15 \\
        Fan & 78.16 & 54.84 & 63.06 & 63.98 \\
        Gearbox & 73.01 & 75.01 & 59.39 & 68.39 \\
        Slider & 96.21 & 61.52 & 58.41 & 68.54 \\
        ToyCar & 79.45 & 77.54 & 59.94 & 71.15 \\
        ToyTrain & 88.04 & 48.57 & 57.12 & 60.66 \\
        Valve & 66.26 & 69.83 & 57.02 & 63.90 \\
        \midrule
        Overall harmonic mean & 75.89 & 64.09 & 58.38 & 65.35 \\
        \bottomrule
    \end{tabular*}
\end{table}

\FloatBarrier
\section{Detection under added interference}
\label{supp:noise}
Tables~\ref{tab:appendix_noise_representative}--\ref{tab:appendix_noise_resnet} report overall $H$ (\%) on development-test sections 00--02. Original input is the unmodified recording, including its recorded background. Each original 10-s recording is mixed with either a normal test recording of a different machine type or Gaussian white noise. The SNR in dB compares original-recording power with added-interference power. For a given recording and interference type, the same waveform is used across systems and scaled to reach each of the seven SNRs. The original normal reference banks, trained networks, embedding-correction statistics and cosine 1-NN rule are held fixed.

RA denotes regional aggregation and EC denotes embedding correction. CMGAN uses two blocks and 64 channels, BEATs uses iter3, the AE uses 128 hidden units and an eight-dimensional bottleneck, and each ResNet uses 18 layers. The fixed BEATs readout averages over time and concatenates frequency positions. Panel (a) in each table uses other-machine operating sounds and panel (b) uses Gaussian white noise. The original-input column reports a separate evaluation of the same checkpoints on the unmodified recordings.

\begin{table}[!htbp]
    \centering
    \caption{Overall $H$ (\%) at each SNR for representative systems and correction configurations.}
    \label{tab:appendix_noise_representative}
    \footnotesize
    \linespread{1.0}\selectfont
    \renewcommand{\arraystretch}{1.12}
    \setlength{\tabcolsep}{2pt}
    \begin{tabularx}{\linewidth}{@{}>{\raggedright\arraybackslash}Xrrrrrrrr@{}}
        \multicolumn{9}{@{}l@{}}{\textbf{(a) Other-machine noise}} \\
        \addlinespace[3pt]
        \toprule
        \textbf{System} & \shortstack{\textbf{Original}\\\textbf{input}} & \textbf{$-9$} & \textbf{$-6$} & \textbf{$-3$} & \textbf{0} & \textbf{$+3$} & \textbf{$+6$} & \textbf{$+9$} \\
        \midrule
        CMGAN + GAP & 58.15 & 49.15 & 49.01 & 49.64 & 50.33 & 52.08 & 53.71 & 55.64 \\
        CMGAN + GAP + EC & 59.88 & 49.28 & 51.29 & 53.41 & 54.10 & 55.45 & 56.93 & 57.98 \\
        CMGAN + RA & 61.85 & 52.59 & 55.48 & 58.01 & 59.79 & 61.02 & 61.31 & 61.72 \\
        CMGAN + RA + EC & 64.20 & 54.64 & 57.51 & 60.70 & 62.58 & 63.66 & 64.21 & 64.43 \\
        BEATs (fixed) & 62.61 & 50.49 & 50.92 & 51.99 & 53.18 & 54.66 & 56.67 & 58.62 \\
        BEATs + RA & 63.76 & 53.98 & 56.21 & 58.74 & 60.60 & 62.09 & 62.56 & 62.93 \\
        BEATs + RA + EC & 65.35 & 55.20 & 56.99 & 59.98 & 62.32 & 63.19 & 63.79 & 64.26 \\
        Direct paired-mixture & 63.59 & 55.93 & 59.54 & 62.70 & 63.67 & 64.32 & 64.10 & 63.91 \\
        Direct paired-mixture + EC & 63.10 & 56.05 & 59.79 & 62.12 & 63.06 & 63.73 & 63.41 & 63.30 \\
        Dis-multi-PL & 65.03 & 52.79 & 53.19 & 54.72 & 56.40 & 57.72 & 59.51 & 61.47 \\
        \bottomrule
    \end{tabularx}
    \par\vspace{0.8em}
    \begin{tabularx}{\linewidth}{@{}>{\raggedright\arraybackslash}Xrrrrrrrr@{}}
        \multicolumn{9}{@{}l@{}}{\textbf{(b) Gaussian white noise}} \\
        \addlinespace[3pt]
        \toprule
        \textbf{System} & \shortstack{\textbf{Original}\\\textbf{input}} & \textbf{$-9$} & \textbf{$-6$} & \textbf{$-3$} & \textbf{0} & \textbf{$+3$} & \textbf{$+6$} & \textbf{$+9$} \\
        \midrule
        CMGAN + GAP & 58.15 & 46.51 & 47.84 & 47.45 & 48.60 & 49.05 & 51.39 & 51.91 \\
        CMGAN + GAP + EC & 59.88 & 46.99 & 46.29 & 46.87 & 46.66 & 48.61 & 50.94 & 53.32 \\
        CMGAN + RA & 61.85 & 48.91 & 54.42 & 56.49 & 58.58 & 60.41 & 60.83 & 61.61 \\
        CMGAN + RA + EC & 64.20 & 43.66 & 51.89 & 55.56 & 59.70 & 62.08 & 63.25 & 64.04 \\
        BEATs (fixed) & 62.61 & 51.33 & 51.02 & 50.39 & 50.55 & 52.01 & 53.23 & 54.88 \\
        BEATs + RA & 63.76 & 51.55 & 53.68 & 59.10 & 60.77 & 61.59 & 61.98 & 62.47 \\
        BEATs + RA + EC & 65.35 & 49.70 & 53.07 & 60.86 & 61.96 & 62.06 & 62.94 & 64.02 \\
        Direct paired-mixture & 63.59 & 53.35 & 57.48 & 61.62 & 63.96 & 64.70 & 64.85 & 64.09 \\
        Direct paired-mixture + EC & 63.10 & 52.49 & 57.37 & 61.44 & 63.38 & 63.84 & 63.97 & 63.71 \\
        Dis-multi-PL & 65.03 & 51.98 & 53.69 & 53.64 & 55.61 & 56.94 & 58.85 & 60.57 \\
        \bottomrule
    \end{tabularx}
\end{table}

\FloatBarrier
\begin{table}[!htbp]
    \centering
    \caption{Overall $H$ (\%) at each SNR for autoencoder and additional pre-trained encoders.}
    \label{tab:appendix_noise_ae_pretrained}
    \footnotesize
    \linespread{1.0}\selectfont
    \renewcommand{\arraystretch}{1.12}
    \setlength{\tabcolsep}{2pt}
    \begin{tabularx}{\linewidth}{@{}>{\raggedright\arraybackslash}Xrrrrrrrr@{}}
        \multicolumn{9}{@{}l@{}}{\textbf{(a) Other-machine noise}} \\
        \addlinespace[3pt]
        \toprule
        \textbf{System} & \shortstack{\textbf{Original}\\\textbf{input}} & \textbf{$-9$} & \textbf{$-6$} & \textbf{$-3$} & \textbf{0} & \textbf{$+3$} & \textbf{$+6$} & \textbf{$+9$} \\
        \midrule
        AE latent & 54.40 & 50.54 & 50.68 & 50.49 & 50.30 & 50.57 & 52.07 & 52.15 \\
        AE residual & 57.66 & 49.48 & 49.34 & 50.44 & 50.62 & 51.18 & 52.58 & 53.88 \\
        AE + RA & 56.35 & 50.39 & 51.35 & 52.78 & 53.82 & 54.90 & 55.04 & 55.47 \\
        AE + RA + EC & 56.93 & 50.57 & 51.94 & 52.91 & 53.55 & 54.93 & 55.92 & 56.65 \\
        CED-base (fixed) & 57.68 & 50.35 & 50.62 & 51.30 & 52.33 & 52.96 & 53.47 & 54.17 \\
        CED-base + RA & 63.70 & 55.55 & 57.66 & 59.74 & 60.90 & 62.04 & 62.34 & 62.90 \\
        CED-base + RA + EC & 61.36 & 55.51 & 57.24 & 59.00 & 59.84 & 60.66 & 60.98 & 61.01 \\
        EAT-base (patch mean) & 58.08 & 50.75 & 51.20 & 51.68 & 52.05 & 52.83 & 54.12 & 55.55 \\
        EAT-base + RA & 60.22 & 53.18 & 54.33 & 56.01 & 57.30 & 58.70 & 59.53 & 59.97 \\
        EAT-base + RA + EC & 61.44 & 53.15 & 55.16 & 56.58 & 57.80 & 58.68 & 59.55 & 60.43 \\
        \bottomrule
    \end{tabularx}
    \par\vspace{0.8em}
    \begin{tabularx}{\linewidth}{@{}>{\raggedright\arraybackslash}Xrrrrrrrr@{}}
        \multicolumn{9}{@{}l@{}}{\textbf{(b) Gaussian white noise}} \\
        \addlinespace[3pt]
        \toprule
        \textbf{System} & \shortstack{\textbf{Original}\\\textbf{input}} & \textbf{$-9$} & \textbf{$-6$} & \textbf{$-3$} & \textbf{0} & \textbf{$+3$} & \textbf{$+6$} & \textbf{$+9$} \\
        \midrule
        AE latent & 54.40 & 50.34 & 51.72 & 49.95 & 48.03 & 47.51 & 49.86 & 49.14 \\
        AE residual & 57.66 & 47.00 & 46.85 & 48.21 & 49.70 & 49.51 & 50.07 & 49.27 \\
        AE + RA & 56.35 & 49.11 & 49.67 & 49.99 & 50.92 & 53.45 & 53.34 & 53.76 \\
        AE + RA + EC & 56.93 & 48.57 & 48.18 & 49.13 & 50.68 & 51.98 & 52.41 & 52.27 \\
        CED-base (fixed) & 57.68 & 52.75 & 51.48 & 51.52 & 51.70 & 52.96 & 54.10 & 54.90 \\
        CED-base + RA & 63.70 & 52.03 & 55.81 & 59.56 & 59.65 & 60.72 & 60.49 & 61.18 \\
        CED-base + RA + EC & 61.36 & 46.20 & 51.71 & 58.68 & 60.66 & 61.17 & 60.47 & 59.98 \\
        EAT-base (patch mean) & 58.08 & 52.58 & 49.27 & 49.98 & 51.02 & 51.15 & 52.00 & 53.73 \\
        EAT-base + RA & 60.22 & 53.47 & 54.56 & 57.02 & 56.66 & 56.42 & 56.81 & 58.22 \\
        EAT-base + RA + EC & 61.44 & 48.00 & 51.08 & 56.51 & 58.16 & 58.45 & 59.21 & 60.08 \\
        \bottomrule
    \end{tabularx}
\end{table}

\FloatBarrier
\begin{table}[!htbp]
    \centering
    \caption{Overall $H$ (\%) at each SNR for ResNet-18 encoders trained with different objectives.}
    \label{tab:appendix_noise_resnet}
    \footnotesize
    \linespread{1.0}\selectfont
    \renewcommand{\arraystretch}{1.12}
    \setlength{\tabcolsep}{2pt}
    \begin{tabularx}{\linewidth}{@{}>{\raggedright\arraybackslash}Xrrrrrrrr@{}}
        \multicolumn{9}{@{}l@{}}{\textbf{(a) Other-machine noise}} \\
        \addlinespace[3pt]
        \toprule
        \textbf{System} & \shortstack{\textbf{Original}\\\textbf{input}} & \textbf{$-9$} & \textbf{$-6$} & \textbf{$-3$} & \textbf{0} & \textbf{$+3$} & \textbf{$+6$} & \textbf{$+9$} \\
        \midrule
        CE & 58.88 & 51.47 & 51.77 & 52.20 & 52.93 & 53.90 & 55.16 & 56.29 \\
        CE + RA & 60.04 & 52.80 & 54.73 & 55.90 & 56.22 & 57.16 & 57.71 & 58.47 \\
        CE + RA + EC & 59.77 & 52.34 & 53.80 & 54.68 & 55.40 & 56.28 & 57.43 & 57.98 \\
        ArcFace & 57.64 & 52.32 & 53.32 & 53.23 & 53.51 & 54.20 & 55.22 & 56.43 \\
        ArcFace + RA & 58.86 & 52.12 & 53.58 & 54.05 & 55.78 & 56.44 & 57.26 & 57.43 \\
        ArcFace + RA + EC & 57.01 & 51.10 & 52.44 & 53.51 & 53.75 & 54.54 & 55.51 & 55.93 \\
        SimCLR & 54.65 & 51.19 & 51.24 & 51.31 & 51.33 & 51.79 & 52.40 & 53.31 \\
        SimCLR + RA & 57.96 & 51.14 & 53.48 & 55.00 & 55.22 & 55.23 & 55.48 & 55.84 \\
        SimCLR + RA + EC & 57.77 & 49.80 & 51.15 & 52.34 & 53.84 & 54.79 & 55.52 & 55.96 \\
        SimSiam & 54.02 & 50.18 & 50.13 & 50.25 & 50.32 & 51.00 & 52.81 & 54.35 \\
        SimSiam + RA & 52.78 & 49.49 & 49.31 & 50.29 & 51.02 & 51.21 & 52.02 & 51.79 \\
        SimSiam + RA + EC & 55.23 & 49.40 & 49.59 & 50.27 & 51.35 & 52.89 & 53.56 & 54.07 \\
        \bottomrule
    \end{tabularx}
    \par\vspace{0.8em}
    \begin{tabularx}{\linewidth}{@{}>{\raggedright\arraybackslash}Xrrrrrrrr@{}}
        \multicolumn{9}{@{}l@{}}{\textbf{(b) Gaussian white noise}} \\
        \addlinespace[3pt]
        \toprule
        \textbf{System} & \shortstack{\textbf{Original}\\\textbf{input}} & \textbf{$-9$} & \textbf{$-6$} & \textbf{$-3$} & \textbf{0} & \textbf{$+3$} & \textbf{$+6$} & \textbf{$+9$} \\
        \midrule
        CE & 58.88 & 51.78 & 53.37 & 55.04 & 54.37 & 56.17 & 58.55 & 59.43 \\
        CE + RA & 60.04 & 52.31 & 50.75 & 51.57 & 56.44 & 57.71 & 58.34 & 58.68 \\
        CE + RA + EC & 59.77 & 50.93 & 51.28 & 51.45 & 56.43 & 56.86 & 57.47 & 57.88 \\
        ArcFace & 57.64 & 54.02 & 55.82 & 56.72 & 56.00 & 53.86 & 54.89 & 57.30 \\
        ArcFace + RA & 58.86 & 52.73 & 53.93 & 55.12 & 57.15 & 58.83 & 58.92 & 58.59 \\
        ArcFace + RA + EC & 57.01 & 48.93 & 51.38 & 49.73 & 50.75 & 55.05 & 57.38 & 56.63 \\
        SimCLR & 54.65 & 49.70 & 48.68 & 47.16 & 48.76 & 50.38 & 51.07 & 52.34 \\
        SimCLR + RA & 57.96 & 49.44 & 51.10 & 52.98 & 54.39 & 55.23 & 56.00 & 56.60 \\
        SimCLR + RA + EC & 57.77 & 48.05 & 48.32 & 50.95 & 52.84 & 54.54 & 55.17 & 55.64 \\
        SimSiam & 54.02 & 49.01 & 49.67 & 49.05 & 48.97 & 49.69 & 49.64 & 51.29 \\
        SimSiam + RA & 52.78 & 49.44 & 50.02 & 50.04 & 49.24 & 48.79 & 48.81 & 50.36 \\
        SimSiam + RA + EC & 55.23 & 49.01 & 48.56 & 49.82 & 49.79 & 49.63 & 51.10 & 52.65 \\
        \bottomrule
    \end{tabularx}
\end{table}